\documentclass[aps,floatfix,onecolumn,showpacs]{revtex4-2}

\usepackage{graphicx}% Include figure files
\usepackage{dcolumn}% Align table columns on decimal point
\usepackage{bm}% bold math
\usepackage{color}

\newcommand{\ve}[1][K]{\mathbf{#1}}

\newcommand{\dd}{\mbox{d}}

\usepackage{amsmath}
\usepackage{amssymb}
\usepackage{latexsym}
\usepackage{multirow}
\usepackage{xcolor}
\usepackage{hyperref}
 
\usepackage{textcomp}

\begin{document}

\title{How stickiness can speed up diffusion in confined systems}

\author{A. Alexandre}
\affiliation{Laboratoire Ondes et Matière d'Aquitaine, CNRS/University of Bordeaux, F-33400 Talence, France}
\author{ M. Mangeat}
\affiliation{Center for Biophysics and Department for Theoretical Physics, Saarland University, D-66123 Saarbr{\"u}cken, Germany}
\author{T. Gu\'erin}
\affiliation{Laboratoire Ondes et Matière d'Aquitaine, CNRS/University of Bordeaux, F-33400 Talence, France}
\author{D.S. Dean} 
\affiliation{Laboratoire Ondes et Matière d'Aquitaine, CNRS/University of Bordeaux, F-33400 Talence, France}
\affiliation{Team MONC, INRIA Bordeaux Sud Ouest, CNRS UMR 5251, Bordeaux INP, Univ. Bordeaux, F-33400, Talence, France}

\bibliographystyle{naturemag}

\date{\today}

\begin{abstract} 
The paradigmatic model for heterogeneous media used in diffusion studies  is built from reflecting obstacles and surfaces. It is well known that the crowding effect produced by these reflecting surfaces slows the dispersion of  Brownian tracers. Here, using a general adsorption desorption model with surface diffusion, we show analytically that making surfaces or obstacles  attractive can accelerate dispersion. In particular, we show that this enhancement of diffusion can  exist even when the surface diffusion constant is smaller than that in the bulk. Even more remarkably, this enhancement effect occurs when the effective diffusion constant, when  restricted to surfaces only, is lower than the effective  diffusivity   with purely reflecting boundaries. We give analytical formulas for this intriguing effect   in periodic arrays of spheres as well as undulating micro-channels. Our results are confirmed by numerical calculations and Monte Carlo simulations. 
\end{abstract}

\maketitle

Determining the transport properties of tracer particles in heterogeneous media at large time and length scales has applications in a range of physical problems including fluid mechanics, hydrology, chemical engineering, soft matter and solid state physics \cite{marbach2018transport,aminian2016boundaries,kim2019tuning,brenner2013macrotransport}. The effective diffusivity %, \textcolor{blue}{which quantifies the spreading dynamics of a cloud of tracer particles,} 
is a crucial input for  problems of mixing \cite{leBorgne2013stretching,dentz2011mixing,barros2012flow}, sorting \cite{bernate2012stochastic}, chemical delivery \cite{aminian2016boundaries,Aminian2015}   as well as chemical reactions \cite{dentz2011mixing,brenner2013macrotransport}. Spatial variations of diffusion and advection  can lead to drastic changes in the effective diffusion constant with respect to homogenous systems. Classic examples include  Taylor dispersion in hydrodynamic flows \cite{taylor1953dispersion}, and the decrease of dispersion due to the energy barriers created by time independent potentials \cite{dean2007effective}. In a number of systems, such as porous media \cite{bhattacharjee2019bacterial}, zeolites, and biological channels \cite{carusela2021computational}, the motion of tracer particles is hindered by hard (technically speaking reflecting) boundaries or obstacles. The effective trapping of the tracer in this case is of  entropic origin but it again leads to a reduction of late time diffusivity \cite{malgaretti2013entropic,burada2009diffusion,jac1967,reguera2006entropic,rubi2019entropic}. The effect of the  confining geometry  on  effective diffusivity has been widely studied \cite{yang2017hydrodynamic,martens2013hydrodynamically,aminian2016boundaries,jac1967,kalinay2006corrections,kalinay2017nonscaling,martens2011entropic,zwanzig1992diffusion,mangeat2017dispersion,mangeat2017geometry,mangeat2018dispersion}, but the vast majority of existing theories focuses on perfectly \textit{reflecting boundaries}. 

\begin{figure}[h!]     
    \includegraphics[width=10cm]{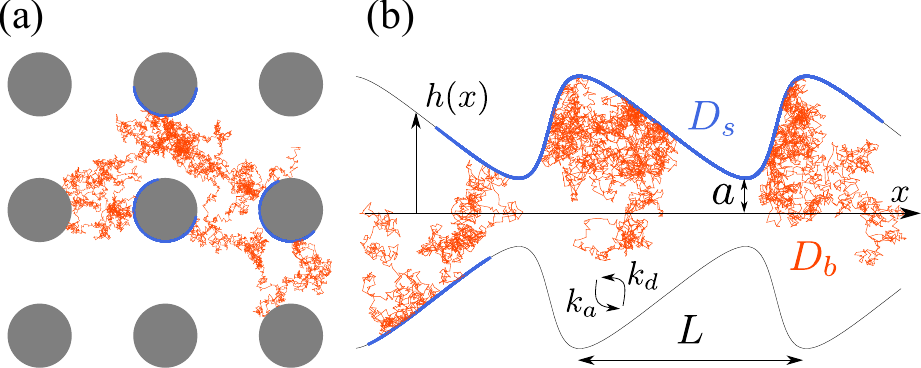}    
    \caption{(color online) Examples of  trajectories  of a tracer particle diffusing in the bulk (orange lines) or on the surface (thick blue lines) in a regular array of spherical obstacles (a) or in a micro-channel (b). }
    \label{schema_figure}
\end{figure}

However, the diffusion of a finite size tracer near a surface is actually much more  complicated than that of simple Brownian motion at a reflecting wall. 
First, the tracer will typically be subject to  non-specific interactions with 
 the surface,  for example as the result of Van der Waals long range attraction and electrostatic potentials, that can be attractive or repulsive, depending on geometry and charges \cite{kim2019tuning,israelachvili2011intermolecular}. Secondly, the components of the diffusion tensor in the vicinity of the wall are reduced due to the no-slip boundary condition on the ambient fluid~\cite{brenner1961slow}. A simple minimal model for these effects is that of a tracer particle that can transiently attach  (or adsorb) to, diffuse on, and detach (desorb) from the surface, thus alternating between phases of bulk diffusion or surface diffusion (Fig. \ref{schema_figure}). This type of model was first proposed in the 90's when it was realized that the effective lateral dispersion of molecules adsorbed at a fluid-solid interface is significantly modified by temporary excursions into  the fluid \cite{bychuk1995anomalous}.  The stochastic motion resulting from  interplay of  bulk and surface diffusion has recently been directly observed experimentally \cite{walder2011single,skaug2013intermittent,chee2016desorption,wang2020non,morrin2020polyelectrolyte}. 

It is also known that surface mediated transport gives rise to non-trivial effects in the context of target search kinetics \cite{benichou2010optimal,calandre2014accelerating,rupprecht2012exact,monserud2016interfacial} and that in certain cases the search efficiency can be tuned via the parameters of the surface mediated diffusion model. A classic related  example is the optimization of  search on DNA by alternating phases of one dimensional diffusion along the DNA, on which the target site is found,  and three-dimensional bulk excursions \cite{von1989facilitated,coppey2004kinetics,berg1981diffusion}. 
It is less clear if similar optimization effects appear for transport properties. 
In Ref.~\cite{putzel2014nonmonotonic}, the effect of adding a short range attraction  to otherwise hard spheres was studied numerically. Remarkably, it was found that spending time trapped at the surface can  lead to an increase in the late time diffusion of the tracer. 
The study of \cite{putzel2014nonmonotonic} neglects the fact that diffusion near the hard sphere is slowed down and one is naturally lead to ask the question as to whether the enhancement effect   persists when this is taken into account. 
Moreover, for general geometries, there are no theories predicting this enhancement of dispersion: existing approaches deal with uniform channels (where the effect is absent) \cite{levesque2012taylor,berezhkovskii2013aris} or in fast exchange limit for non-planar geometries \cite{quintard1994convection,brenner2013macrotransport,edwards1993dispersion}, where no increase of dispersion due to sticky surfaces was found. 

%\textcolor{green}{Clearly if the attractive potential is too strong the particle will spend a long time trapped at the surface of individual spheres and diffusion will ultimately be reduced. This suggests that there is an optimal trapping potential to increase diffusion. } 
%A theoretical formalism to analyze   surface mediated transport  has already been used to predict dispersion in the limit of fast exchange between bulk and surface, but no increase of dispersion due to sticky surfaces was found. Effective dispersion in presence of surface mediated diffusion has been studied in \textit{uniform} channels , however the  effect of increased diffusivity is absent in this particular geometry. 

In this Letter, we introduce a theory that quantitatively predicts the increase in late time dispersion (with respect to its value for reflecting boundaries) induced by making the  boundaries attractive, even when the surface diffusivity is lower than the microscopic bulk diffusivity. %We  present a general formalism to compute the late time diffusion constant  in arbitrary periodic media with surface mediated transport. %We show that the effects seen in \cite{putzel2014nonmonotonic}  can occur generically in surface mediated diffusion models and that they can be much more pronounced, even when the reduction of diffusion near the surface is taken into account. 
As an example, we give  analytical formulas characterizing dispersion in non-dilute spherical sticky obstacles, which explain the simulation results  of \cite{putzel2014nonmonotonic}. Results are then given for slowly undulating channels, in  this case the increase of dispersion turns out to be stronger. %In both cases, the increase of dispersion due to weakly attractive surfaces is generic and can even occur  for slow surface diffusion. 
For strongly undulating channels, we show that enhancement  of dispersion no longer occurs. Our results identify in which situations dispersion  is optimized by the interplay between diffusion in bulk and on the surface. The effect uncovered here can be viewed as an example of {\em catalysis} for diffusion where the introduction of the {\em surface state} via reaction with the surface induces an increase in the rate of dispersion.

%%%%%%%%%%%%%%%%%%%%%%%%%%%%%%%%%%%%%%%%%%

\noindent\textit{Model of surface mediated transport - }We consider a tracer particle, of position $\ve[r](t)$ at time $t$ in a $d$-dimensional space,  an heterogeneous medium with obstacles, or confining boundaries. The particle can either diffuse in the \textit{bulk} (``b"), with a bulk molecular diffusivity $D_b$, and local drift $\ve[u]_b(\ve[r])$, or diffuse along the \textit{surface} (``s'') of the obstacles or confining surface, where it diffuses with diffusion coefficient $D_s$ (typically smaller than $D_b$).   In our approach, we assume that the tracer particle is point-like. This assumption can be made without loss of generality since the problem of a finite size tracer particle can be studied by modifying the obstacle geometry. In this way, the particle size influences the effective diffusivity. We will also neglect inertial effects, which are irrelevant for sufficiently small tracers in viscous fluids. We can  also consider a local drift field $\ve[u]_s(\ve[r])$, within the surface. By $k_d$ we denote a detachment rate at which the tracer desorbs from the surface. When the  transition  between bulk and surface is viewed as a reaction, the binding kinetics  is quantified by an imperfect reactivity parameter $k_a$ \cite{grebenkov2019imperfect}, which has the dimension of a velocity. In the context of Taylor dispersion it has been   shown how the parameters $k_a$ and $k_d$ can be determined from a microscopic model \cite{AlexandreGeneralized2021}. 

The probability densities $p_b(\ve[r],t)$ and $p_s(\ve[r],t)$ to be at position $\ve[r]$ at time $t$ in the bulk and surface    obey
\begin{align}
 \frac{\partial p_b}{\partial t}    &= \nabla\cdot [D_b\nabla  p_b - \textbf{u}_b(\ve[r]) p_b ], \label{FKPbulk}\\%\equiv \mathcal{L}_B \ p_b \\
\frac{\partial p_s }{\partial t}  & = \nabla_{s}\cdot [D_s \nabla_s   p_s - \textbf{u}_s(\ve[r]) p_s ] - k_d \ p_s  +k_a \ p_b ,  \label{FKPSurf}
\end{align}
where $\nabla_{s}$ is the surface  nabla  operator. The boundary conditions, determined from probability conservation, are given by
\begin{equation}
\textbf{n}\cdot \left[D_b  \nabla p_b  - \textbf{u}_b(\textbf{r}) p_b   \right]= k_d\ p_s  -k_a \ p_b,  \label{boundary}
\end{equation}
where $\textbf{n}$ is the surface normal vector, pointing away from the bulk. Finally, we assume that the geometry of the medium, as well as all fields $\ve[u]_b(\ve[r]),\ve[u]_s(\ve[r])$, are spatially periodic.

\noindent\textit{Effective transport properties - } The late time transport properties in a given spatial direction (say the $x$ direction) are quantified by an effective drift $v_e$ and effective diffusivity $D_e$ in this direction:
\begin{align}
v_e = \frac{\overline{ x(t)-x(0)}} {t}, \ D_e = \lim_{t\to\infty}\frac{\overline{ [x(t)-x(0)-v_e t]^2}}{2t},
\end{align} 
where the overline denotes the  average over stochastic trajectories. The average drift is straightforward to calculate (see Supplementary Information (SI)) 
\begin{align}
&v_e= \int_{V} d\ve[r]  \ (\nabla x) \cdot \ve[j]_b  \ +\int_{S} dS \ (\nabla_s x) \cdot \ve[j]_s, \label{Eqve} \\
&\ve[j]_b=(\ve[u]_b -D_b \nabla)  P_{b}^{st},\ \  \ve[j]_s=(\ve[u]_s  -D_s \nabla_s ) P_{s}^{st},
\end{align}
where the integrals are evaluated over the volume $V$ and the surface $S$ of the walls in one period. The functions $P_{b}^{st}(\ve[r])$ and $P_{s}^{st}(\ve[r])$  are the stationary probability densities of the position modulo the period and  are  time-independent solutions of Eqs.~(\ref{FKPbulk}) and (\ref{FKPSurf}) with periodic conditions. %\textcolor{red}{Note that when the surface is not flat, $\nabla_s x$ is spatially dependent.} 

To compute $D_e$ in the absence of any advection, we use the fluctuation dissipation 
theorem that relates it to the effective drift when a small force $F$ acts on the tracer particle:
\begin{align}
D_e(F=0)= k_BT \times \left[\frac{d}{dF}\ v_e(F) \right]_{F=0}, \label{FluctuationDiss}
\end{align}
where $k_BT$ is the thermal energy. Here, a force $F$ in the $x$ direction corresponds to the following  drift fields:  
\begin{align}
\ve[u]_b(\ve[r])=\beta D_b F (\nabla x), \ \ve[u]_s(\ve[r])=\beta D_s F (\nabla_s x),
\end{align}
 with  $\beta=1/k_BT$. %This above remark Eq. (\ref{FluctuationDiss}) means that the dispersion results from a perturbative analysis of the effective drift in the limit of low force. 
 At zero force, we denote the stationary probability densities by $P_{b}^{st,0},P_{s}^{st,0}$, both are  uniform  and are given by: 
 \begin{align}
 P_{b}^{st,0}(\ve[r])=\frac{1}{V+S\delta}, \ P_{s}^{st,0}(\ve[r]) = \delta P_{b}^{st,0}, \label{Pstat}
 \end{align}
where $\delta =k_a/k_d$  is an adsorption length (that can be much larger than the range of interactions with the surface). The length $\delta$ quantifies how ``sticky'' the surface is, with surfaces becoming non-reflecting in our problem when $\delta$ is comparable to $V/S$, in which case the fraction of time spent  on the surface become significant.
We now introduce two auxiliary fields $f_{b}$ and $f_{s}$ which quantify the deviation of the stationary probability density from the uniform distribution at low forces: 
\begin{align}
    P_{b}^{st} \simeq P_{b}^{st,0}+\beta F f_{b}(\ve[r]), \ P_{s}^{st} \simeq P_{s}^{st,0}+\beta F f_{s}(\ve[r]).  \label{LinearAnsatz}
\end{align}
  Inserting this ansatz into Eqs.~(\ref{Eqve}) and (\ref{FluctuationDiss}), we see that $D_e(F=0)$  (denoted simply by $D_e$)  is given by
  \begin{align}
      D_e   =  &D_b  \int_{V} \dd \textbf{r} \ (\nabla x) \cdot \nabla \left[ P_{b}^{st,0}    x  -     f_{b} \right]  \nonumber \\
 & +    D_s \int_{S} \dd S \  (\nabla_s x) \cdot \nabla_s \left[   P_{s}^{st,0}    x     -   f_{s}  \right].
       \label{deff_kubo}
\end{align}
The equations for  $f_{b}, f_{s}$ are  obtained by inserting the ansatz (\ref{LinearAnsatz}) into the transport equations (\ref{FKPbulk}, \ref{FKPSurf},  \ref{boundary}) and expanding to linear order in $F$, which yields
\begin{align}
     &    \nabla^2   f_{b} = 0, \hspace{0.5cm}
          D_s          \nabla_{s}^2  [f_{s}- P_{s}^{st,0}x]     =k_d f_{s}  -k_a f_{b},\label{Eqf}\\
    &D_b   \  \textbf{n}\cdot   \nabla    f_{b}  = k_d f_{s} - k_a f_{b} + D_b P_{b}^{st,0}\ (\nabla x)\cdot \textbf{n}.
    \label{Eqboundary}
\end{align}
The above equations, together with periodic boundary conditions, define $f_b$ and $f_s$ up to an unimportant additive constant.   This kind of   formula linking  micro and macro-transport properties can also be  found  in macrotransport theory \cite{brenner2013macrotransport}, homogenization methods  or Kubo formulas \cite{Guerin2015Kubo,guerin2015}. In the case of surface mediated transport, the above equations are general and  in the limit of fast exchange between bulk and surface ($k_d,k_a\to\infty$ keeping $k_a/k_d$ constant) reproduce the results given in \cite{quintard1994convection,brenner2013macrotransport,edwards1993dispersion}. 
Note that surface curvature effects arise in Eq.~(\ref{Eqf}) via the term $\nabla_s^2x$ which does not vanish (as it would in the bulk). The above partial differential equations can be numerically solved by standard finite element routines. However we will also derive   analytical results  in certain limits. 
 %In the very special case of channels of uniform width, the above equations are straightforward to solve, leading to the trivial result
%$D_e=  (|\Omega|D_b +|\Sigma|\delta D_s )/(|\Omega|+|\Sigma|\delta)$.
%In this case, $D_e$ is simply the average of the surface and bulk diffusivities. However, for non-trivial geometries, we show below that this property is no longer valid. 

\begin{figure}[h!]
    \includegraphics[width=8cm]{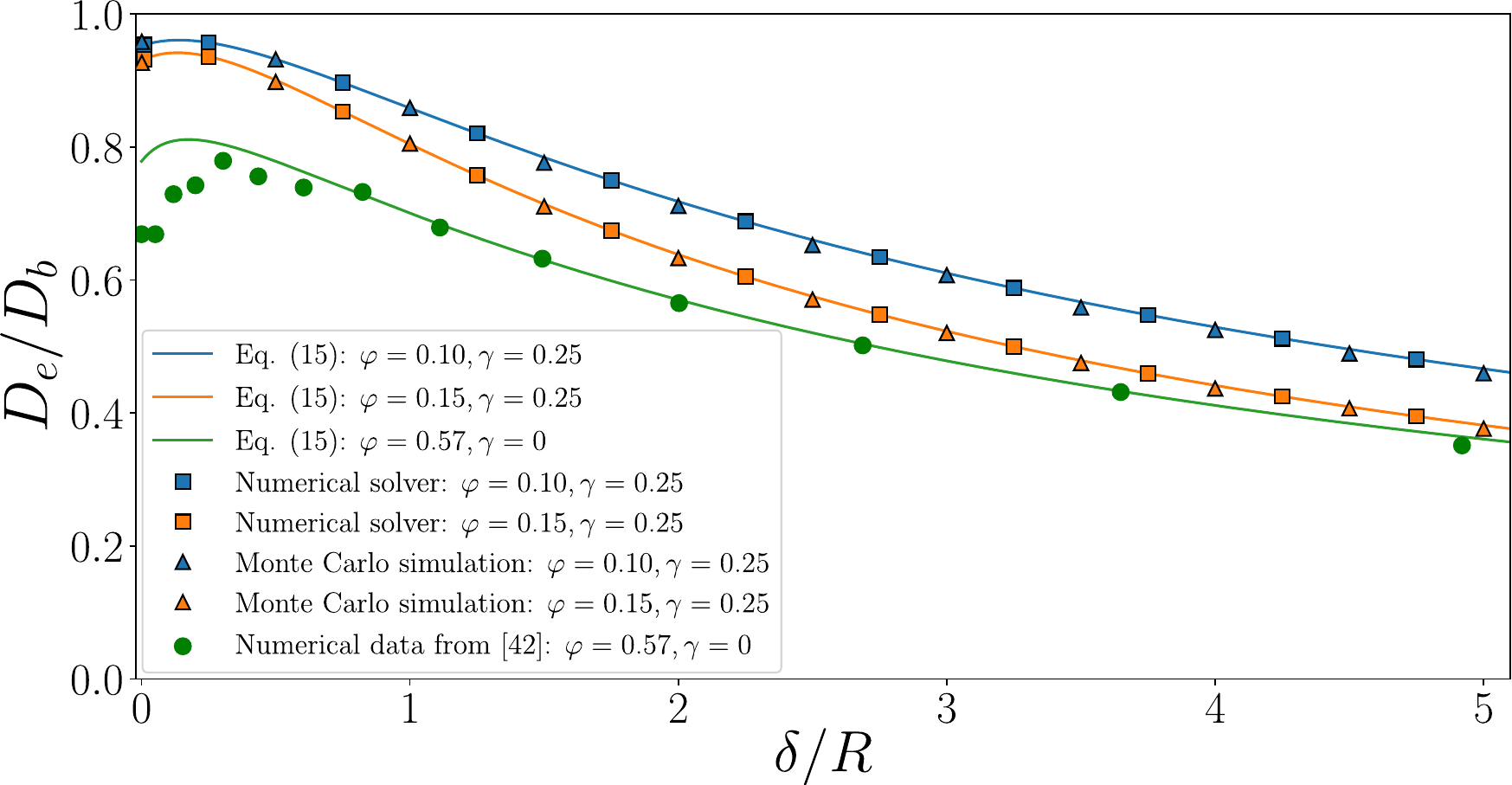}
      \caption{(color online) Long-time dispersion for a periodic array of spherical obstacles ($d=3$) with $D_s = D_b$. Lines: theoretical prediction, Eq. (\ref{D_e_obstacles}). Squares: data obtained by numerical integration of  Eqs.~(\ref{Eqf}), (\ref{Eqboundary}). Triangles: results of stochastic simulations of the stochastic process $\ve[r](t)$ associated to the Fokker-Planck equations (\ref{FKPbulk},\ref{FKPSurf}). Circles: Monte Carlo simulation data taken from \cite{putzel2014nonmonotonic}, where the tracer particles are submitted to an  exponentially  decaying potential. The parameter $\delta$ was adjusted so that the stationary probability densities (\ref{Pstat}) in our description are the same as in the case of this potential (see SI for details).} 
    \label{fig_deff_obstacles_main}
\end{figure}
  
\textit{Dispersion in regular arrays of spheres - } 
One the most studied models  of a crowded environment is that of a regular array of spherical obstacles. Here we consider the cases of a square lattice (in 2D, $d=2$) or a cubic lattice (in 3D, $d=3$), see Fig. \ref{schema_figure}(a). We denote by $L$ the distance between the centers of nearest neighbouring spheres, $R$ their radius, and $\varphi$ their volume fraction. An exact analytical expression for the dispersion for any $\varphi$ is difficult to obtain even without the surface interactions. However, in the limit of small $R/L$ (equivalently, the small $\varphi$ limit), we can look for solutions in terms of matched asymptotic expansions, with an inner solution for $f_{b}(\ve[r])$ that varies at the scale $\vert \ve[r]\vert\sim R$ , and an outer solution  varying at the scale $L$.   The auxiliary fields in the limit $R\to0$ are written as
\begin{align}
f_{b}(\ve[r])=\begin{cases}
\sum\limits_{n\ge0}   R^{n+1} f_{b}^{(n)}(r/R,\omega)   & \left(r \ll L, \text{inner} \right)  \\
 \sum\limits_{n\ge0}  R^{d+n} F_{b}^{(n)}(r,\omega)   & (r \gg R, \text{outer} )
\end{cases},
\label{AnsatzSPheres}
\end{align} 
where $r$ is the distance to the center of the nearest obstacle and $\omega$ denote angular variables. %Here, the first line describes the behavior of the solution in the limit $R\to 0$ at $r/L$ constant, while the second line describes the behavior of $f_{\Omega}$ when $R\to0$ at fixed position $r$. 
%The relevant orders of $R$ are in fact identified from the calculation. 
The solutions at successive orders are found by inserting Eq.~(\ref{AnsatzSPheres}) into our formalism and imposing that both outer and inner solutions  are equal in the regime $R\ll r\ll L$ (where they are both valid). To  order $\varphi^4$ (see SI), we find
\begin{align}
D_e=&\frac{D_b}{1+(d\xi-1)\varphi}\frac{1-\alpha \varphi}{1+\alpha \varphi/(d-1)} +o(\varphi^{4}) 
\nonumber,\\
\alpha=&\frac{D_b +D_s\xi(\gamma-1)(d-1)}{D_b+D_s \xi [ \gamma(d-1)+1]}, \ \gamma=\frac{D_b}{k_a R}, \ \xi=\frac{\delta}{R}.\label{D_e_obstacles}
\end{align}
The above formula is exact up to the order $\varphi^4$   in the small $\varphi$ limit, when $\xi $ and $\gamma$ are kept constant. It is  valid for a  large range of values of $\varphi$, as shown in Fig. \ref{fig_deff_obstacles} where it is compared to numerical solution of the full transport equations (\ref{Eqf}, \ref{Eqboundary}).  The theory also agrees with the Monte Carlo simulations of Ref.~\cite{putzel2014nonmonotonic}.  This expression (\ref{D_e_obstacles}) is a generalization of similar Maxwell type formulas obtained for reflecting obstacles, see \textit{e.g.} Refs.~ \cite{mangeat2020effective,kalnin2002calculations,MaxwellBook,lebenhaft1979diffusion}. We can define a critical surface diffusivity $D_s^*$ as the value of $D_s$ above which a weak attraction increases $D_e$ (so $\left[\partial D_e/\partial \delta \right]_ {\delta =0} =0$ at $D_s=D_s^*$). Using Eq.~(\ref{D_e_obstacles}), we find 
\begin{align}
D_s^*=   D_b\left(1-\frac{1-\varphi}{d} \right).  
\end{align}
Interestingly, $D_s^*$ depends only on $\varphi$ and is independent of the adsorption and desorption rates. At this level of perturbation theory we also see that   $D_s^*<D_b$. 
Consequently, if the surface and bulk diffusivities are equal ($D_s=D_b$), making the surfaces weakly attractive \textit{generically increases}  dispersion. This effect holds even when $D_s<D_b$ (as long as $D_s> D_s^*$). 
Hence, spending time on a surface, even  with reduced  diffusion, can actually lead to faster dispersion  than in the case of purely reflecting obstacles. 
 A similar formula to Eq. (\ref{D_e_obstacles}) was found in Ref. \cite{edward1995diffusion} in the fast exchange limit ($\gamma=0$), however this formula does not predict an increase of dispersion for weakly attractive spheres, and it disagrees with our theory and with numerical simulations  (see SI). Finally, the  increase in $D_e$   is rather marginal in this geometry, as can be seen in Fig.~\ref{fig_deff_obstacles}.

\begin{figure}[h!]
    \includegraphics[width=8cm]{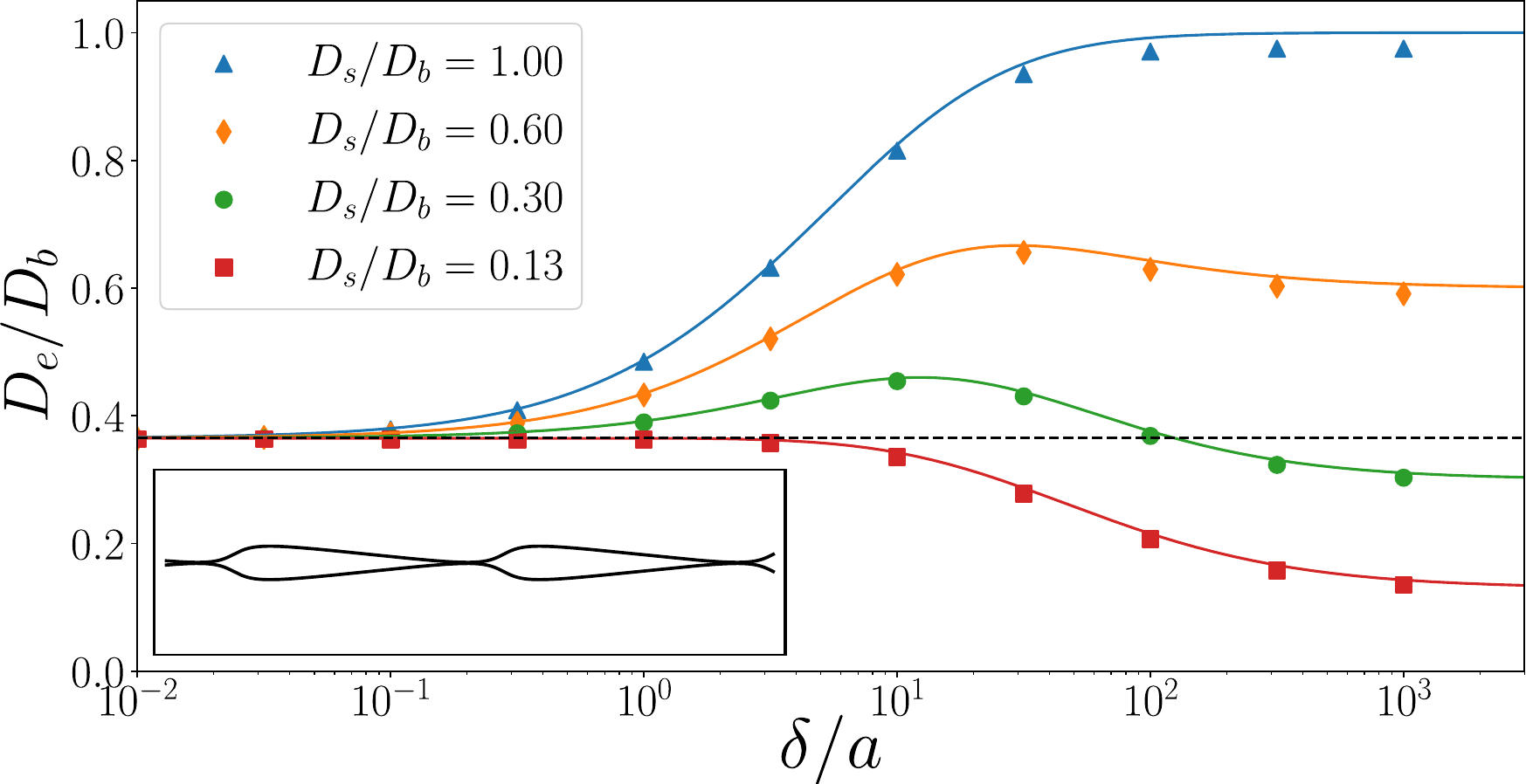}
    \caption{(color online)  Long-time diffusion coefficient for a two-dimensional channel of height profile $h/a =  16+ 15\arctan[\cos u /(\sin u + 3/2)]/\arctan(2/\sqrt{5}) $ with $u=2\pi x/L$. Symbols: results of the numerical integration of Eqs.~(\ref{Eqf}) and (\ref{Eqboundary}) for $k_a a /D_b = 10$  and $L/a=500 $ ; for this value the channel shape is shown in the inset. Lines: predictions of Eq. (\ref{D_LJ_D_no_flow}) in the limit of slowly varying width. }
    \label{fig_deff_channel}
\end{figure}

\noindent\textit{Dispersion in slowly undulating channels - } We now analyze the case of a surface-mediated diffusion process in a symmetric channel of half-width $h(x)$ (if $d=2$) or in an axisymmetric channel of radius $h(x)$ (if $d=3$), see figure \ref{schema_figure}(b). 
Such geometries are often viewed as a paradigm for  transport in confined media \cite{malgaretti2013entropic,reguera2006entropic}. 
We consider the limit of slowly varying undulations ($L\to\infty$), where the problem can be solved using a perturbation expansion:
\begin{align}
f_{b} =\sum_{n\ge0} \frac{1}{L^{n}} f_{b}^{(n)}\left(\frac{x}{L},\ve[r]_\perp \right),  f_{s} =\sum_{n\ge0}   \frac{1}{L^{n}} f_{s}^{(n)}\left(\frac{x}{L}\right),
\end{align}
where $\ve[r]_\perp$ is the position perpendicular to the direction $x$. Inserting the above ansatz into our formalism and writing all equations order by order, we obtain (see SI)
\begin{align}
D_e \underset{L\to\infty}{=} \frac{1}{\left\langle h^{d-2}\left( h + (d-1)\delta \right)  \right\rangle   \left\langle \frac{h^{2-d}}{D_b h+ (d-1)D_s \delta }  \right\rangle},
\label{D_LJ_D_no_flow}
\end{align}
where $\left< \cdot\right> = \frac{1}{L} \int_0^{L} \cdot\ dx $ denotes the spatial average over one period $L$. Note that $D_e$ is not given by a simple steady state average between effective bulk diffusion and surface diffusion as it is the case for a flat channel \cite{levesque2012taylor}. For reflecting walls ($\delta=0$), we recover the well known result obtained with the Fick-Jacobs approximation, $D_e(\delta=0)=D_b[\langle h^{d-1}\rangle\langle h^{-(d-1)}\rangle]^{-1}$ \cite{jac1967, zwanzig1992diffusion}. 
%\textcolor{red}{As opposed to the system of arrays of spheres, the system here will exhibit a non-zero diffusion constant even when the particle is strongly confined at the wall (the limit $\delta\to\infty$), here we see that $D_e(\infty)=D_s[\langle h^{d-2}\rangle\langle h^{-(d-2)}\rangle]^{-1}$.}
Once again, defining $D_s^*$ as the value of $D_s$ for which $[\partial D_e/\partial\delta]_{\delta=0}=0$, we find  
\begin{align}
D_s^*= D_b \frac{\langle h^{d-2} \rangle\langle h^{1-d} \rangle}{\langle h^{d-1} \rangle\langle h^{-d} \rangle}.  
\end{align}
It can be shown that $D_s^*$ is   smaller than $D_b$ for any choice of profile $h$. Figure \ref{fig_deff_channel} illustrates the increase of dispersion when surfaces are made weakly attractive for one example of a two-dimensional channel. Interestingly, the increase of dispersion can be as high as $250\%$  (upper curve), this change could  be made arbitrarily high for sharper channel undulations. The enhancement effect can thus be significantly larger than that observed for  sticky spheres as studied above and in Ref.~\cite{putzel2014nonmonotonic}. 

It is clear that the enhancement of diffusion is related to the fact that diffusing along the surface avoids bottlenecks (or entropic barriers) in the channel. However, the effect of the surface interaction is more subtle than just this basic effect. 
As seen in Figure \ref{fig_deff_channel} from the curve with solid circles, a finite interaction with the surface can enhance diffusion with respect to both the  case $\delta=0$ (pure reflection) and $\delta\to\infty$ (diffusion restricted to the surface), even when the surface diffusion is $30\%$ of that of bulk diffusion. 
Actually, it can be shown that when $D_e(0)=D_e(\infty)$, then the inequality $D_s>D_s^*$ always holds (see SI). Hence, for any channel geometry, there is  a regime for which the effective diffusivity is larger than the effective diffusivity for completely reflecting ($\delta=0$) or completely sticky ($\delta\to\infty$) boundaries. The subtlety of the effect is also highlighted by the earlier results on spherical obstacles, where the enhancement effect actually shows up at first order in the volume fraction, thus in absence of significant bottlenecks. 

%The forms of $D_s^*$, $D_e(0)$ and $D_e(\infty)$ can be used to establish some interesting bounds between these quantities and are discussed in the SI.

\textit{Dispersion in highly undulating channels -} Finally, we investigate  rapidly varying channels ($L\ll a$, with $a$ the minimal channel height), for which we obtain (see SI) 
\begin{align}
D_e&\underset{L\ll a}{=}   \frac{ V_c   }{V + S\delta } D_b, \label{DeStrongCorr}%\\&= \frac{a^{d-1}}{\langle h^{d-2}\left(h+(d-1)\delta\sqrt{1+h'^2} \right) \rangle}
\end{align}
where $V_c$ is the volume of a central region (a cylinder of radius $a$, so $V_c=2aL$ in 2D and $V_c=\pi a^2 L$ in 3D). This formula can be deduced from a simple ergodic argument. In the central region, the tracer particle can freely diffuse, but in the peripheral regions  made of very thin dead-ends, the tracer particle does not contribute to dispersion along the channel axis (at leading order). Hence, the late time diffusivity is the product of $D_b$ with the fraction of time spent in the central region, leading to Eq.~(\ref{DeStrongCorr}). 
In this geometric limit, we see that surface interactions only \textit{decrease} dispersion.

\noindent\textit{Conclusion - } It is  well established that first passage kinetics to a target can be optimized by an appropriate combination of sejourns of diffusion in spaces of different dimensions \cite{calandre2014accelerating,coppey2004kinetics,benichou2010optimal}, as exemplified by the search of a gene sequence on DNA \cite{von1989facilitated,berg1981diffusion,coppey2004kinetics}. However, this optimization is not linked to an acceleration of diffusivity. Here, we have theoretically demonstrated a similar optimization effect holds for transport features: dispersion in crowded media can be enhanced if the obstacles or surfaces exhibit a short ranged attraction for the tracer.  
 This enhancement phenomenon was first noted in Ref.~\cite{putzel2014nonmonotonic} for hard spheres. Using a surface mediated diffusion model, we have developed a general transport theory to analyze the effect of short range attractive interactions on dispersion. This theory explains the results of Ref.~\cite{putzel2014nonmonotonic} and we have provided analytical results for dilute systems (which  work rather well even for large volume fractions).  
We have also analyzed dispersion in symmetric channels where the particle can adsorb and detach from the walls. For slowly undulating channels, we have shown that an even stronger  diffusion enhancement can occur and have given exact results. In both geometries, we find that even when the surface diffusion  constant $D_s$ is smaller than that in the bulk $D_b$ (as must be the case physically) this enhancement can still occur and we have determined the critical values of $D_s$ above which this happens. 
  
Computer time for this study was provided by the computing facilities MCIA (Mesocentre de Calcul Intensif Aquitain) of the Universit\'e de Bordeaux and of the Universit\'e de Pau et des Pays de l’Adour.
%\newpage 
  
  \vspace{2cm}

\section*{Supplementary Information}
%\center{\textbf{Supplemental material}}
  
%\vspace{1cm}
  
Here, we provide details of calculations that support the results of the main text:
\begin{itemize}
\item a proof of the general expression of the average velocity $v_e$ (Section \ref{CalculationDrift}),
\item some reminders of tensor analysis  for diffusion on surfaces (Section \ref{MetricSection}),
\item a description of stochastic simulations of the process $\ve[r](t)$ (Section \ref{SectionSimulations})
\item a derivation of the expression of $D_e$ in periodic arrays of spherical obstacles, including a comparison with the results of the literature (Section \ref{SectionSpherres}),
\item a derivation of the dispersion for slowly varying and fast varying channel widths (Sections \ref{SlowChannel} and \ref{FastChannel}). 
\end{itemize}

\appendix
\section{Calculation of the effective drift - derivation of Eq. (5)}
\label{CalculationDrift}
Here, we compute the effective drift defined in Eq.~(4). For now, we consider a fixed initial position $\ve[r]_0$ and  initial state $i_0$ (equal to $b$ or $s$, i.e. in the bulk or on the surface). The displacement in the $x$ direction after a time $t$ reads
 \begin{equation}
    \overline{x(t) - x_0} =  \int_{\Omega} d\mathbf{r}     (x-x_0) p_b( \mathbf{r},t| \ve[r]_0,i_0) + \int_{\Sigma} dS(\mathbf{r}) (x-x_0)p_s( \mathbf{r},t |\ve[r]_0,i_0)~,    
    %\int_{B} d\mathbf{y} P_{b}^{st}(\mathbf{y}) g_{B}( \mathbf{y},t) + \int_{S} dS(\mathbf{y}) P_{s}^{st}(\mathbf{y}) g_S( \mathbf{y},t),
\end{equation}
where $\Omega$ and $\Sigma$ are  the volume and   surface of the whole  medium. Taking the temporal derivative and using Eqs.~(1,2) yields
\begin{align}
\partial_t \overline{x(t)}  = & \int_{\Omega} d\mathbf{r}\  
    (x-x_0) \nabla \cdot \left[ (D_b \nabla - \mathbf{u}_b)  p_b( \mathbf{r} ,t|\mathbf{r}_0,i_0) \right] \nonumber\\
    &+ \int_{\Sigma} dS(\mathbf{r}) (x-x_0)\left\{\nabla_s \cdot  \left[( D_s \nabla_s - \mathbf{u}_s )p_s( \mathbf{x} ,t |\mathbf{r}_0,i_0)\right] - k_d \ p_s( \mathbf{r} ,t|\mathbf{r}_0,i_0) + k_a \ p_b( \mathbf{r} ,t | \mathbf{r}_0,i_0) \right\}~. 
\end{align}
After performing integrations by parts and using the boundary condition Eq.~(3), we obtain
\begin{align}
\partial_t \overline{x(t)} = \int_{\Omega} d\mathbf{x} (\nabla x) \cdot  \left(\mathbf{u}_b -D_b \nabla \right) p_b(\mathbf{r} ,t | \mathbf{r}_0,i_0) +\int_{\Sigma} dS(\mathbf{x}) (\nabla_s x) \cdot \left(\mathbf{u}_s -D_s \nabla_s \right) p_s(\mathbf{r}  ,t | \mathbf{r}_0,i_0) ~.\label{Eq9531}
\end{align}
We now introduce  $P_s(\ve[r],t)$ and $P_b(\ve[r],t)$, which are the probability densities at $t$ to be respectively on the surface or in the bulk, at position $\ve[r]$ \textit{modulo the period}, \textit{i.e.}:
\begin{align}
    P_{b}(\mathbf{r},t) = \sum_{\ve[L]} p_b(\mathbf{r} + \ve[L] ,t)~, \ P_{s}(\mathbf{r},t) = \sum_{\ve[L]} p_s(\mathbf{r} + \ve[L],t)~,
\end{align}
where  the vectors $\ve[L]$  are all the  vectors such that the environments at $\ve[r]$ and $\ve[r]+\ve[L]$ are identical (for example, in the channel geometry $\ve[L]=n L\ve[e]_x$ with $n$ integer). In Eq.~(\ref{Eq9531}),  $\nabla x$, $\nabla_s x$, $\ve[u]_b$ and $\ve[u]_s$ are periodic in $\ve[r]$. Therefore, splitting the integral into multiple integrals (each of them covering one period) and using the above relation yields
\begin{align}
\partial_t \overline{x(t)} = \int_{V} d\mathbf{r} \ (\nabla x)\cdot \left(\mathbf{u}_b -D_b \nabla \right) P_b(\mathbf{r} ,t | \mathbf{r}_0,i_0) +\int_{S} dS(\mathbf{r})\  (\nabla_s x)\cdot \left(\mathbf{u}_s -D_s \nabla_s \right) P_s(\mathbf{r}  ,t | \mathbf{r}_0,i_0)~, \label{0542}
\end{align}
where $V$ and $S$ are the volume and surface in one period of the channel only. This result is obtained by assuming a fixed initial state and position, let us now extend it to the case that $\ve[r]_0$ and $i_0$ are drawn from the stationary distribution of the dynamics modulo the period.  In this case, it is useful to note that 
\begin{equation}
   P_i^{st}(\mathbf{r} ) = \int_{V} d \mathbf{r}_0 P_i(\mathbf{r} ,t | \mathbf{r}_0,b) P_{b}^{st}(\mathbf{r}_0)
   +\int_{S} dS(\mathbf{r}_0) P_i(\mathbf{r} ,t | \mathbf{r}_0,s) P_{s}^{st}(\mathbf{r}_0) \ \ ~\forall t~, \hspace{2cm}  (i\in \{b,s\})
\end{equation}
so that averaging Eq.~(\ref{0542}) over initial conditions and integrating over $t$ leads to
\begin{align}
v_e  =\partial_t \overline{x(t)}  =\frac{\overline{x(t)-x(0)}}{t}= \int_{V} d\mathbf{r} \ (\nabla x) \cdot \left(\mathbf{u}_b -D_b \nabla \right) P_b^{st}(\mathbf{r}) +\int_{S} dS(\mathbf{r})\  (\nabla_s x)\cdot \left(\mathbf{u}_s -D_s \nabla_s \right) P_s^{st}(\mathbf{r})~,   
\end{align}
which is Eq.~(5) in the main text. Note that this expression also holds, at long times only, for any choice of initial condition, as can be seen by taking the long time limit of Eq.~(\ref{0542}).

\section{Tensor analysis for surface diffusion}
\label{MetricSection}
We recall some  basic results of tensor analysis that are useful to interpret explicitly $\nabla_s^2x$ in different coordinate systems. We assume that the surface in a $d$ dimensional space is parametrized by the coordinates $y^i$ ($1\leq i \leq d-1$). If  $\ve[r]_s$ is the position vector on the surface, then  the metric given  by
\begin{equation}
    g_{ij}=\textbf{e}_{s,i}\cdot \textbf{e}_{s,j}~,    \hspace{1cm}\textbf{e}_{s,i}=\frac{\partial \textbf{r}_s}{\partial y^{i}}~,
\end{equation}
where the suffix $s$ indicates that the position remains on the surface. We use the notation $|g|=\det(g_{ij})$ and $g^{ij}$ for the  inverse of the metric tensor (i.e. $g^{ij}g_{jk}=\delta_{ik}$, with Einstein's summation convention). In our formalism, we need to identify the elementary surface elements, the surface gradient and Laplacian of a scalar function $\phi$, which are well known and are  given by
\begin{equation}
    dS=dy^{1}dy^{2}...dy^{d-1}\sqrt{|g|}~, \hspace{1cm} \nabla_s \phi =    g^{ij}\ \frac{\partial \phi}{\partial y^{j}}\ \textbf{e}_{s,i}~,   \hspace{1cm}  \nabla_s^{2}\phi = \frac{1}{\sqrt{|g|}}\frac{\partial}{\partial y^{i}} \left(\sqrt{|g|}\ g^{ij}\ \frac{\partial \phi}{\partial y^{j}}\right)~.
\end{equation}

\section{Algorithm for stochastic simulations}
\label{SectionSimulations}
We have used the following algorithm to simulate the stochastic process $\ve[r](t)$ associated with the Fokker-Planck equations (1),(2)   in periodic arrays of spheres (the algorithm could be adapted to other geometries):
\begin{itemize}
\item a particle in the bulk  evolves with $dr_i=\sqrt{2D_{b}}dW_i$ (with $\langle dW_i\rangle=0$ and $\langle dW_idW_j\rangle=dt\delta_{ij}$) during the time step $dt$ in each direction $i=\{x,y,z\}$. 
\item At the end of a step, if the tracer particle ends up  inside the obstacle, it switches to the attached state with probability $P_{\text{adsorb}} = k_{a} \sqrt{\pi dt/D_{b}}  $ and  it is reflected with probability $1-P_{\text{adsorb}}$, see Ref.~\cite{singer2008partially}.
\item An attached particle switches to the detached state with probability  $ 1 - e^{-k_{d} dt}$.
\item If the particle is attached to a spherical  3D obstacle, the stochastic equation (with Ito convention) for the angular coordinates $(\theta,\phi)$ is
\begin{equation}
    d \theta_t =  \frac{D_{s} \ dt}{R^2 \tan \theta_t}+ \frac{\sqrt{2 D_{s}}}{R} dW_{\theta} \ , \ 
    d \phi_t =   \sqrt{\frac{2 D_{s}}{R^2 \sin^2\theta_t}} dW_{\phi} ~,  \  \left<dW_{\phi}^2\right> =\left< dW_{\theta}^2 \right> = dt.  
    \label{langevin_Surface}
\end{equation}
%The algorithm can be adapted for other geometries.%, for example at the surface of a 2D channIf it is attached at the surface of a 2D channel, it evolves with 
%\begin{align}
%dx_t= - \frac{ D_s h'(x_t) h''(x_t) dt}{(1+ h'^2(x_t))^2} + \sqrt{ \frac{2 D_s dt}{1+ h'^2(x_t)}}  dW_{x},\  dW_x^2 = dt.
%\end{align}
 This equation can be read off from the Fokker-Planck equation for $p_s(\ve[r],t)$, which in terms of spherical angles $(\theta,\phi)$ reads (in absence of detachment event) :
\begin{align}
\partial_t p_s(\theta,\phi,t)=\frac{D_s}{R^2\sin\theta}\partial_\theta[\sin\theta \partial_\theta p_s]+\frac{D_s}{R^2\sin^2\theta}\partial_\phi^2 p_s.
\end{align}
Here, $p_s dS=p_s R^2\sin\theta d\theta d\phi$ is the probability to find the particle with the angular coordinates within the interval $[\theta,\theta+d\theta]\times[\phi,\phi+d\phi]$. We now consider $q(\theta,\phi)$ the probability distribution function of $(\theta,\phi)$ so that $q(\theta,\phi)d\theta d\phi$ is the probability to observe the particle in  $[\theta,\theta+d\theta]\times[\phi,\phi+d\phi]$. We thus have $q =p_s R^2\sin\theta$ and $q$ obeys 
 %\begin{equation}
%	\frac{\partial p}{\partial t} = \frac{D_{\text{s}}}{R^2} \left[ \frac{1}{\sin^2 \theta} \frac{\partial^2 p }{\partial \phi^2} +  \frac{1}{\sin \theta} \frac{\partial }{\partial \theta} \left( \sin  \theta \frac{\partial p}{\partial \theta} \right)  \right] 
%\end{equation}
%with the normalization $\int_{0}^{2 \pi } \dd \phi \int_{0}^{\pi} \dd \theta R^2 \sin \theta\  p(\theta, \phi) = 1$. Now, we consider the density $q$ such that $q(\theta , \phi) \dd \theta \dd \phi = p(\theta , \phi) R^2 \sin \theta\dd \theta \dd \phi$. Its equation satisfies:
\begin{equation}
	\frac{\partial q}{\partial t} = \frac{\partial^2 }{\partial \phi^2} \left[ \frac{D_{\text{s}}}{R^2 \sin^2 \theta} q \right] + \frac{\partial }{\partial \theta } \left[ \frac{\partial}{\partial \theta } \left( \frac{D_{\text{s}}}{R^2} q \right) - \frac{D_{\text{s}}}{R^2 \tan \theta }q\right].
\end{equation}
It is clear that the stochastic differential equation that corresponds to this Fokker-Planck equation is Eq.~(\ref{langevin_Surface}).
\end{itemize}

\section{Dispersion in periodic array of spheres - derivation of Eq. (15)}
\label{SectionSpherres}

\subsection{2D obstacles}

Here, we compute $D_e$ for a  square array of 2D spheres of radius $R$.  The volume fraction is given by  $\varphi = \pi R^2/L^2$.
We chose the units of length so that $L=1$, without loss of generality. We  use polar coordinates $(r,\theta)$, with $r$ the distance to the center of the disk and $\theta$ the angle with the $x$-axis. The surface is parametrized by the coordinate $y^1=\theta$. Using the formulas of Section \ref{MetricSection},  we obtain 
$\nabla_s^2 x  =-\cos\theta/R$. Equations (12, 13) written in polar coordinates are: 
\begin{equation}\left\{
    \begin{split}
        &\text{(bulk) } \frac{1}{ r }\frac{\partial}{\partial  r }\left(r \frac{\partial f_{b}}{\partial r }\right) + \frac{1}{r ^{2}}\frac{\partial^{2}f_{b}}{\partial \theta ^{2}} = 0 ~,& (r>R)\\
        &\text{(surface) } \frac{D_s}{R^2}  \frac{\partial^{2}f_{s} }{\partial \theta ^{2}} 
        = k_d f_s(\theta)-k_a f_b(r=R,\theta) - D_s  P_{s}^{st,0} \frac{\cos \theta}{R} ~, & (r=R)\\
        &\text{(boundary condition) } - D_b\frac{\partial f_{b}}{\partial r} (r=R,\theta) = k_d f_{s} - k_a f_{b} - D_b     P_{b}^{st,0} \cos \theta ~. & (r=R)
    \end{split}\right. \label{Equations2DDisks}
\end{equation}
 with $P_{b}^{st,0}=1/(1+ (2\delta /R-1) \varphi)$ and $P_{s}^{st,0} = \delta P_{b}^{st,0}$.  After an integration by parts in Eq.~(11), we find that  the  effective diffusion constant is 
\begin{equation} 
    D_{e} = D_{b}(1-\varphi) P_{b}^{st,0} + D_s \varphi \xi P_{b}^{st,0} + D_{b} R\int_{0}^{2\pi}  \dd \theta \ \cos(\theta)f_{b}(r=R,\theta)- D_{s}\int_{0}^{2\pi}\dd \theta \cos(\theta) f_{s}(\theta)~.
\label{deff_obstacles_2D}
\end{equation}
We now introduce  the dimensionless parameters $\xi=\delta/R$ and $\gamma=D_b/(k_a R)$. 
We wish to compute $D_e$ in the limit $R\to0$ (or, equivalently, the limit of vanishing volume fraction $\varphi \to0$), when  $\xi$ and $\gamma$ are kept constant. We solve this problem by using strongly localized perturbation analysis \cite{ward1993}, considering that the solution $f_b$ varies at two different length scales $r\sim R$ and $r \sim 1$ when $R\to 0$ (we recall that the period is $L=1$). We  introduce an \textit{inner solution} (valid near the obstacles), depending on $\tilde{r}=r/R$ and $\theta$, which has an expansion in the limit $R\rightarrow 0$  at fixed $\tilde{r}=r/R$:
\begin{equation}
%\left\{
    \begin{split}   
        f_{b}(\ve[r])  = P_{b}^{st,0} R\left[ f_{b }^{0}(\tilde{r},\theta) + R^{2} f_{b }^{1}(\tilde{r},\theta) +R^{4}  f_{b }^{2}(\tilde{r},\theta) + ...\ \right],    
        f_{s}(\ve[r])  = P_{b}^{st,0} R^{2}\left[ f_{s }^{0}(\theta) + R^{2} f_{s}^{1}(\theta) +R^{4}  f_{s}^{2}(\theta) + ...\ \right].
        \end{split}
%     \right.
%    \hspace{1cm}(R\to0, \ \text{fixed }\  \tilde{r}=r/L,\theta)
     \label{Ansatz043}
\end{equation}
Note that we have introduced the factor $P_{b}^{st,0}$ since it is obvious in Eq.~(\ref{Equations2DDisks}) that the solutions are proportional to this factor. We introduce an \textit{outer solution}, far from the obstacle, which admits the following expansion when $R\rightarrow 0$ at fixed $r$:
\begin{equation}
    f_{b}(\ve[r]) = P_{b}^{st,0} R^{2}\left[ F_{b }^{0}(r,\theta) + R^{2} F_{b}^{1}(r,\theta) +R^{4}  F_{b}^{2}(r,\theta) + ...\ \right].\label{AnsatzOuter}
%    \hspace{1cm} (R\rightarrow 0, \ \text{fixed } \ r,\theta). 
\end{equation}
It follows that all $f_b^i(\tilde{r},\theta)$ and $F_b^i(r,\theta)$ are harmonic functions for all orders $i$, and that all $F_b^i(r,\theta)$ must satisfy periodic conditions on the sides of the unit square. The equations for $f_s^i$ and for the value of $f_b^i $ near the obstacles are found by inserting Eq.~(\ref{Ansatz043})  into Eq.~(\ref{Equations2DDisks}) and identifying terms order by order, leading to: 
\begin{equation} 
         D_s\gamma\xi\frac{\partial^{2}f_{s}^{i}}{\partial \theta ^{2}} 
        = D_b\left( f_{s}^{i} -\xi f_{b}^{i}\right) - \delta_{i,0} \xi^{2}D_s \gamma  \cos \theta, \hspace{1cm}
          -\xi\gamma \frac{\partial f_{b}^{i}}{\partial \tilde{r}} (\tilde{r}=1,\theta) = f_{s}^{i} - \xi f_{b}^{i} - \delta_{i,0}\xi \gamma \cos \theta,
\label{eq_inner_solution}
\end{equation}
 where $\delta_{i,0}$ is   the Kronecker delta symbol. Additional conditions to determine  $f_b^i $ and $F_b^i$ come from the requirement that the inner and the outer solution coincide in their regime of common validity  $R\ll r\ll 1$, so that the expansion of Eq.~(\ref{Ansatz043}) for $\tilde{r}\to\infty$ coincides with the expansion of (\ref{AnsatzOuter}) for $r\to0$. %\tr{To do so, we match iteratively each terms of the inner  solution with the outer solution, so that both perturbative expansions in $R$ are identical.}
%\textcolor{blue}{OR: More explicitly, if there is a term $f_b^i(\tilde{r},\theta)= A \tilde{r}^n+...$ for the expansion of $f_b^i $ at large $\tilde{r}$, this means that $f_b\sim A R^{1+2i+n} r^n $ for the expansion of $f_b$ in the intermediate region ($R\ll r\ll 1$), imposing that $F_b^{2i+n-1}(r,\theta)$ displays the same term $Ar^n$ in its expansion near $r=0$.  The reverse is true, and it turns out that the solutions can be constructed iteratively, first constructing $f_b^0$, then $F_b^0$, then $f_b^1$, followed by $F_b^1$, $f_b^2$, and so on. }\textcolor{green}{[David please decide between blue and red version ]}

At leading order, $f_b^0$ is the harmonic function that does not diverge for large $\tilde{r}$ and that satisfies Eq.~(\ref{eq_inner_solution}) for $i=0$:
\begin{align}
    f_{b}^{0}(\tilde{r}, \theta )  = \underbrace{\frac{A_{0}}{\tilde{r}}  \cos(\theta)}_{\text{must match } F_{b}^{0}},~ f_{s }^{0}( \theta ) =\tilde{f}_{s}^{0} \cos(\theta),~  
     A_{0} =- \alpha,~  
     \hspace{0.3cm}\tilde{f}_{s}^{0} = \xi   \ \frac{D_s \xi (\gamma +1)-D_b }{D_s \xi (\gamma +1)+D_b}~,  \hspace{0.3cm} \alpha=  \frac{D_b+D_s \xi (\gamma -1)}{D_b+D_s \xi (\gamma +1)}~.\label{85432}
\end{align}
 We see that $f_b^0 \approx  A_0\cos\theta/\tilde{r}$ for large $\tilde{r}$, so that $f_b\simeq A_0 \cos\theta R^2/r$ in the   region $R\ll r\ll1$. This implies that the leading order of the outer solution must be of order $R^2$, as assumed in Eq.~(\ref{AnsatzOuter}), and that  $F_b^0(r\to0,\theta) = A_0 \cos\theta/r$, as indicated in Eq.~(\ref{85432}). Since $F_b^0$ is also harmonic with periodic boundary conditions, it can be expressed as  
\begin{align}
    F_{b}^{0}(r,\theta ) = -2\pi A_{0} \ \textbf{e}_{x}\cdot\nabla G \label{EqFb0}~,
\end{align}    
where $G$ is the pseudo Green's function for the unit-square with periodic  conditions, which satisfies the equation: 
\begin{equation}
    - \nabla^{2}G =\delta(x)\delta(y)-1~,\ G(x\pm1,y)=G(x,y\pm1)=G(x,y)~.
\end{equation}
The Green's function $G$ is given in closed form in Ref.~\cite{mamode2014fundamental}, its behaviour near the origin reads 
\begin{equation}
      G(r,\theta) = -\frac{\log r}{2\pi} +\frac{r^{2}}{4}+C  - \frac{C_0}{8\pi}  r^4\cos(4\theta) + \mathcal{O}(r^8)~,
\end{equation}
with $C_0$ a  constant that we will not need to detail here. Eq.~(\ref{EqFb0}) is justified by noting that $F_{b}^{0}$ near the origin is
\begin{align}
     F_{b}^{0}(r,\theta)= \underbrace{\frac{A_{0}}{r}\cos(\theta)}_{\text{ matching } f_{b}^{0}} \underbrace{-~\pi A_{0}\cos(\theta) r}_{\text{must match } f_{b}^{1}} ~+~\underbrace{C_{0}\cos(3\theta)r^{3}}_{\text{must match } f_{b}^{2}} +~ \mathcal{O}(r^7)~, \label{8542}
\end{align}
where  we see that the $1/r$ term correctly matches the corresponding $1/\tilde{r}$ term in Eq.~(\ref{85432}). We have also indicated that the terms of order $r$ and $r^3$ will have to be matched by corresponding terms $\tilde{r}$ and $\tilde{r}^3$ in the expansions of $f_b^1$ and $f_b^2$, respectively.  In particular,  $f_b^1(\tilde{r},\theta)\simeq -\pi A_{0}\tilde{r}\cos\theta$ for large $\tilde{r}$, since $f_b^1$ is harmonic,  it is of the form
\begin{align}
    f_{b}^{1}(\tilde{r}, \theta) = \underbrace{\frac{A_{1}}{\tilde{r}}\cos(\theta)}_{\text{must match } F_{b}^{1}} +~ \underbrace{B_1\tilde{r}\cos(\theta)}_{\text{matching } F_{b}^{0}}~, \hspace{1cm} B_1=-\pi A_{0}.
\end{align}
 Solving Eq.~(\ref{eq_inner_solution}) for $i=1$, we obtain $f_s^1$ and  the constant $A_1$:
 \begin{align}
    A_{1} =  \alpha B_1 ~, \hspace{1cm} f_{s}^{1}(\theta) =\tilde{f}_{s}^{10} \cos(\theta)~,   \hspace{1cm} \tilde{f}_{s}^{10} =   \frac{2  \xi D_b B_1 }{D_s \xi (\gamma +1)+D_b}~.
\end{align}
The outer solution at next-to-leading order  $F_b^1$ and its expansion near the origin then reads:
\begin{equation}
    F_{b}^{1}(r,\theta ) = -2\pi A_{1} \textbf{e}_{x}\cdot\nabla G = \underbrace{\frac{A_{1}}{r}\cos(\theta)}_{\text{ matching } f_{b}^{1}} \underbrace{-\pi A_{1}\cos(\theta) r}_{\text{must match } f_{b}^{2}} +\underbrace{C_{1}\cos(3\theta)r^{3}}_{\text{must match } f_{b}^{3}}+ \mathcal{O}(r^7)~.\label{843295}
\end{equation}
\textit{Second order:} We continue the iteration and we look for $f_b^2$ in the form 
\begin{equation}
        f_{b}^{2}(\tilde{r},\theta) = \underbrace{\frac{A_{2}}{\tilde{r}}\cos(\theta)}_{\text{must match } F_{b}^{2}} + \underbrace{B_{2}\tilde{r}\cos(\theta)}_{\text{matching }F_{b}^{1}} +\underbrace{\frac{K_{2}}{\tilde{r}^{3}}\cos(3\theta)}_{\text{must match } F_{b}^{3}} + \underbrace{C_0 \tilde{r}^{3}\cos(3\theta)}_{\text{matching } F_{b}^{0}}~, \hspace{1cm} B_{2} = -\pi A_{1} ~,
\end{equation}
so that it matches corresponding terms for $F_{b}^{1}$ and $F_{b}^{0}$ in Eqs.~(\ref{8542}) and (\ref{843295}), respectively. The expression for $A_{2}$ and $  {f}_{s}^{2}$ can be found by solving Eq. (\ref{eq_inner_solution}) with $i=2$:
\begin{equation}
 A_{2} = \alpha B_{2} ~,     \    f_{s}^{2}(\theta)  = \tilde{f}_{s}^{20}\cos(\theta )+\tilde{f}_{s}^{21}\cos(3\theta )~,    \
     \tilde{f}_{s}^{20} =  \frac{2 \xi D_b B_{2} }{D_s \xi (\gamma +1)+D_b}~.
    \label{inner_order2}
\end{equation}
The values of $K_{2}$ and $ \tilde{f}_{s}^{21}$ are not written   because they do not contribute to $D_e$. For the outer solution, we obtain
\begin{equation}
    F_{b}^{2}(r,\theta ) = -2\pi A_{2} \textbf{e}_{x}\cdot\nabla G = \underbrace{\frac{A_{2}}{r}\cos(\theta)}_{\text{ matching } f_{b}^{2}} \underbrace{-\pi A_{2}\cos(\theta) r}_{\text{must match } f_{b}^{3}} +\underbrace{C_{2}\cos(3\theta)r^{3}}_{\text{must match } f_{b}^{4}} +~ \mathcal{O}(r^7)~.
\end{equation}
\textit{Third order:} At this order, the expression of the inner solution reads
\begin{equation}
    f_{b}^{3}(\tilde{r},\theta) = \underbrace{\frac{A_{3}}{\tilde{r}}\cos(\theta)}_{\text{must match } F_{b}^{3}} +~ \underbrace{B_{3}\tilde{r}\cos(\theta)}_{\text{matching } F_{b}^{2}} ~+\underbrace{\frac{K_{3}}{\tilde{r}^{3}}\cos(3\theta)}_{\text{must match } F_{b}^{4}} +~ \underbrace{C_1 \tilde{r}^{3}\cos(3\theta)}_{\text{matching } F_{b}^{1}}~, \hspace{1cm} B_{3} = -\pi A_{2}~.
\end{equation}
This form is chosen so that the corresponding terms   in $F_{b}^{2}$ and $F_{b}^{1}$ are matched. Identically to Eq.~(\ref{inner_order2}), we get
 \begin{equation}
    A_{3} = \alpha B_{3} ~,    \    f_{s}^{3}(\theta)  = \tilde{f}_{s}^{30}\cos(\theta )+\tilde{f}_{s}^{31}\cos(3\theta )~,    \
    \tilde{f}_{s}^{30} =  \frac{2 \xi D_b B_{3} }{D_s \xi (\gamma +1)+D_b} ~.
\end{equation}
At this order of perturbation, we find that the effective diffusivity as given in Eq.~(\ref{deff_obstacles_2D}) is given by:
\begin{align}
        D_{e} =  P_{b}^{st,0} \left\{ D_b(1-\varphi) + D_s\xi \varphi + D_b \varphi \left[ A_{0} + \sum_{i=1}^{3}R^{2i}(A_{i} +B_{i})\right]
        -D_s\varphi \sum_{i=0}^{3}R^{2i}\tilde{f}_{s}^{i} \right\}+o(R^8)~.
\end{align}
An explicit calculation of all terms, followed by a re-summation, leads to the  result for $D_e$ for 2D circular obstacles:
\begin{equation}
         D_{e} = D_b P_{b}^{st,0}\left(1-2\alpha \varphi +2\alpha^2 \varphi^2-2\alpha^3 \varphi^3+2\alpha^4 \varphi^4 \right)+o(\varphi^4)=\frac{D_b}{1+ (2\xi-1)\varphi} \ \frac{1-\alpha \varphi}{1 + \alpha \varphi}
  +o(\varphi^4)  \label{obstacles_2D_final}~.
\end{equation}

\subsection{3D spherical obstacles}
We now consider a 3-dimensional cubic array of spheres of radius $R$. Here, we use  spherical coordinates $(r,\theta, \phi)$, and we  calculate the late time diffusion constant in the $z$ direction, the distance between neighboring spheres is again taken as $L=1$. Note that $r=0$ corresponds to the center of an obstacle, and the nearest obstacles point to the directions $(\theta=\pi/2,\phi=n\pi/2$), and  $\theta=0$ and $\theta=\pi$.  
The long-time  diffusion constant is given by
\begin{equation}
        D_e =P_{b}^{st,0}\left[D_{b}\left(1-\varphi\right) +2 \varphi \xi D_s\right]
        + \int_{0}^{\pi} \dd \theta \int_{0}^{2\pi} \dd \phi  \sin(\theta)\cos(\theta) \left(D_{b}R^2 f_{b}-2D_{s}Rf_s\right)~,
\label{deff_obstacles_3D}
\end{equation}
where $ \varphi= 4\pi R^{3}/3$ is the volume fraction. We adapt the method exposed in the 2D case, by looking for an inner solution, for which the ansatz reads: 
\begin{equation}
\left\{
    \begin{split}   
f_{b}(\ve[r]) &= P_{b}^{st,0} R\left[ f_{b }^{0}(\tilde{r},\theta,\phi) + R f_{b }^{1}(\tilde{r},\theta,\phi) +R^{2}  f_{b }^{2}(\tilde{r},\theta,\phi) + ...\ \right]~,\\
f_{s}(\ve[r])& = P_{b}^{st,0} R^{2}\left[ f_{s}^{0}(\theta,\phi) + R f_{s}^{1}(\theta,\phi) +R^{2}  f_{s}^{2}(\theta,\phi) + ...\ \right]~,
        \end{split}\right.\hspace{1cm}(R\to0, \ \text{fixed }\  \tilde{r}=r/L,\theta,\phi)
\end{equation}
where we have factorized by $P_{b}^{st, 0} = \left[1+(3\xi-1)\varphi\right]^{-1}$. The outer solution is
\begin{equation}
    f_{b}(\ve[r]) = P_{b}^{st,0} R^{3}\left[ F_{b }^{0}(r,\theta,\phi) + R F_{b}^{1}(r,\theta,\phi) +R^{2}  F_{b}^{2}(r,\theta,\phi) + ...\ \right]~. \hspace{1cm}(R\to0, \ \text{fixed }\  r,\theta,\phi) 
\end{equation}
Note that  we have included all powers of $R$ in our ansatz although many of these terms will be found to be zero afterwards. The solutions at successive orders are found by requiring that (i) all $f_b^i$ and $F_b^i$ are harmonic functions, (ii) $F_b^i$ satisfy periodic boundary conditions, (iii) the behavior of the inner and the outer solution is the same at all orders in the region $R\ll r\ll 1$, and (iv) the inner functions satisfy  
\begin{equation}\left\{
    \begin{split}         
        &\text{(surface) }\gamma D_s\xi \left[ \frac{1}{\sin(\theta)}\frac{\partial}{\partial \theta }\left( \sin(\theta) \frac{\partial f_{s}^{i}}{\partial \theta} \right)+ \frac{1}{\sin(\theta)^{2}} \frac{\partial^{2} f_{s}^{i}}{\partial \phi^{2}}\right]
        = D_b\left(f_{s}^{i} -\xi f_{b}^{i}\right) - 2\delta_{i,0} \xi^{2}D_s \gamma  \cos(\theta)~,\\
        &\text{(boundary condition) }-\xi\gamma \frac{\partial f_{b}^{i}}{\partial \tilde{r}} (\tilde{r}=1,\theta) = f_{s}^{i} - \xi f_{b}^{i} -\delta_{i,0}\xi  \cos(\theta)~.
    \end{split}\right.
\label{eq_kubo_3D_obstacles}
\end{equation}

\textit{At leading order}, the inner solution is determined from Eq. (\ref{eq_kubo_3D_obstacles}) and reads
\begin{align}
    f_{b}^{0}(\tilde{r}, \theta ) =\underbrace{\frac{A_{0}}{\tilde{r}^{2}} \cos(\theta)}_{\text{must match } F_{b}^{0}},~  f_{s }^{0}( \theta ) =\tilde{f}_{s}^{0} \cos(\theta),~ A_{0} =-\frac{\alpha}{2},~ \ \alpha= \frac{2 D_s \xi (\gamma -1)+D_b }{D_s \xi (2\gamma +1)+D_b},\
    \tilde{f}_{s}^{0} =  \frac{\xi[2 D_s \xi (2\gamma +1)-D_b] }{2[D_s \xi (2\gamma +1)+D_b]}~.
\end{align}
  It is useful to  introduce the pseudo-Green's function $G$ with periodic boundary conditions,  which satisfies 
\begin{align}
-\nabla^2G=\delta(\ve[r])-1~, \ G(x\pm1,y,z)=G(x,y\pm1,z)=G(x,y,z\pm1)=G(x,y,z).
\end{align}
In fact, we will not need the full expression of $G$ but only its   behavior near the origin, namely 
\begin{align}
 G(r,\theta,\phi) \underset{r\to0}{=} \frac{1}{4\pi r} +\frac{r^{2}}{6}+r^{4}A_{4}(\theta,\phi)+r^{8}A_{8}(\theta,\phi) + ...
\end{align}
with $A_{i}(\theta, \phi)$ functions that are built from Legendre polynomials and respect the symmetry of the problem. For example, $A_4$ is built as a combination $a P_4^0(\cos\theta)+b P_4^4(\cos\theta)\cos(4\phi)$ (with $P_n^m$ the Legendre polynomials) with the requirement that,  for $\phi= n \pi /2$ (with $n$ an integer), $A_4$ does not depend on $\cos(2\theta)$. The only possible form for $A_4$ is 
\begin{equation}
   A_{4}(\theta,\phi) = C_{0}\left\{  20\cos(2\theta)+35\cos(4\theta)+9 - 5 \cos(4\phi) [4\cos(2\theta)-\cos(4\theta)-3]\right\}
   \end{equation}
and  $C_0$ a constant that we will not  need to determine here.  We claim   that  $F_{b}^{0}(r,\theta ) = -4\pi A_{0} \textbf{e}_{z}\cdot\nabla G$. Indeed, this is a harmonic function with periodic boundary conditions, and its expansion near the origin reads:
\begin{equation}
        F_{b}^{0}(r,\theta ) = -4\pi A_{0} \textbf{e}_{z}\cdot\nabla G = \underbrace{\frac{A_{0}}{r^{2}}\cos(\theta)}_{\text{ matching } f_{b}^{0}} \underbrace{-\frac{4\pi}{3} A_{0}\cos(\theta) r}_{\text{must match } f_{b}^{3}} ~+~\underbrace{D_{0}[3\cos(\theta)+5\cos(3\theta)]r^{3}}_{\text{must match } f_{b}^{5}}~+ \underbrace{E_{0}(\theta,\phi)r^{7}}_{\text{must match } f_{b}^{9}}+~ o(r^{7})~,\label{895442}
\end{equation}
with $D_{0}=-128\pi C_0$   and $E_{0}(\theta,\phi)$ a   combination of Legendre polynomials, for which we will only need the   property:
\begin{equation}
\int_{0}^{\pi}\dd \theta\int_{0}^{2\pi}\dd \phi \sin(\theta)\cos(\theta) E_{0}(\theta,\phi) = 0~. \label{legendre}
\end{equation}

Next, at first and second order, we obtain the trivial solutions  $f_b^1=f_s^1=F_b^1=0$ and   $f_b^2=f_s^2=F_b^2=0$. The inner solution at third order is obtained by solving Eq.~(\ref{eq_kubo_3D_obstacles}) with $i=3$ and requiring that the large distance behavior matches with the corresponding term in the expansion of $F_b^0$ in Eq.~(\ref{895442}), we obtain
\begin{equation}
	\begin{split}
	f_{b}^{3}(\tilde{r}, \theta) = \underbrace{\frac{A_{3}}{\tilde{r}^{2}}\cos\theta}_{\text{must match } F_{b}^{3}} + \underbrace{B_{3}\tilde{r}\cos\theta}_{\text{matching } F_{b}^{0}} ~, \
	f_{s}^{3}(\theta) = \cos\theta \tilde{f}_{s}^{3} ~, \\  B_{3} = -\frac{4\pi}{3} A_{0} ~, 
	A_{3} =\frac{B_{3}}{2} \alpha, \ \tilde{f}_{s}^{3} =  \frac{3 \xi D_b B_{3} }{2\left(D_s \xi (2\gamma +1)+D_b \right)}~.
	\end{split}
\end{equation}
The outer solution at this order and its expansion near the origin is
\begin{equation}
        F_{b}^{3}(r,\theta,\phi ) = -4\pi  A_{3} \textbf{e}_{z}\cdot\nabla G = \underbrace{\frac{A_{3}}{r^{2}}\cos(\theta)}_{\text{ matching } f_{b}^{3}} ~\underbrace{-\frac{4\pi}{3} A_{3}\cos(\theta) r}_{\text{must match } f_{b}^{6}} ~+~\underbrace{D_{1}[3\cos(\theta)+5\cos(3\theta)]r^{3}}_{\text{must match } f_{b}^{8}} ~+ ~\mathcal{O}(r^{7})~.
\end{equation}
Next, the inner and outer solutions at fourth order vanish, $f_b^4=f_s^4=F_b^4=0$. The inner solution  $f_{b}^{5}$ reads:
\begin{align}
    f_{b}^{5}(\tilde{r},\theta) =\underbrace{ \left[3\cos(\theta ) +5\cos(3\theta)\right] \frac{A_{5}}{\tilde{r}^{4}}}_{\text{must match } F_{b}^{7}} ~+~ \underbrace{D_0\left[3\cos(\theta ) +5\cos(3\theta)\right]\tilde{r}^{3}}_{\text{matching } F_{b}^{0}} ~, ~ f_{s}^{5}(\theta) = \left[3\cos(\theta ) +5\cos(3\theta)\right] \tilde{f}_{s}^{5}~.
\end{align}
The constants $A_5$ and $\tilde{f}_{s}^{5}$ are found by solving (\ref{eq_kubo_3D_obstacles}) with $i=5$  but their precise expressions are not necessary because 
\begin{equation}
    \int_{0}^{\pi}\dd \theta\int_{0}^{2\pi}\dd \phi \cos(\theta)\sin(\theta)\left[3\cos(\theta ) +5\cos(3\theta)\right]=0~,
\end{equation}
so that this order does not  contribute in the expression of  $D_e$. For the outer solution, there are no terms to be matched so $F_{b}^{5} = 0$.
We continue the iteration and we find at order $6$:
\begin{align}
  &  f_{b}^{6} = \underbrace{\frac{A_{6}}{\tilde{r}^{2}}\cos(\theta)}_{\text{must match } F_{b}^{6}}+~ \underbrace{B_{6}\tilde{r}\cos(\theta)}_{\text{matching } F_{b}^{3}}  ,
    f_{s}^{6}(\theta) = \cos(\theta )\tilde{f}_{s}^{6}, \ B_{6} = -\frac{4\pi}{3} A_{3}~,  A_{6} =\frac{B_{6}}{2} \alpha~, \
    \tilde{f}_{s}^{6} =  \frac{3 \xi D_b B_{6} }{2\left(D_s \xi (2\gamma +1)+D_b\right)}~,\\    
&        F_{b}^{6}(r,\theta,\phi ) = -4\pi A_{6} \textbf{e}_{z}\cdot\nabla G = \underbrace{\frac{A_{6}}{r^{2}}\cos(\theta)}_{\text{ matching } f_{b}^{6}} ~\underbrace{-\frac{4\pi}{3} A_{6}\cos(\theta) r}_{\text{must match } f_{b}^{9}} ~+~\underbrace{D_{2}[3\cos(\theta)+5\cos(3\theta)]r^{3}}_{\text{must match } f_{b}^{11}}~+~ \mathcal{O}(r^{7})~.
\end{align}
 For the inner solution at $7^{th}$ order, there is no term to be matched, so   $f_b^7=f_s^7=0$. However, for the outer solution $F_{b}^{7}$, we have to match a term from $f_{b}^{5}$.  We can build  $F_{b}^{7}$ from $( \ve[e]_z\cdot\nabla)^2  F_b^0 $ so that  its expansion near the origin reads 
\begin{equation}
        F_{b}^{7}(r,\theta) = \frac{4A_{5}}{3A_{0}}( \ve[e]_z\cdot\nabla)^2  F_b^0         =\underbrace{ \frac{A_{5}}{r^{4}}\left[ 3\cos(\theta)+5\cos(3\theta) \right]}_{\text{ matching } f_{b}^{5} }~+~ \underbrace{\frac{64 A_{5}D_{0}}{A_{0}}r\cos(\theta)}_{\text{must match } f_{b}^{10}} ~+~ \mathcal{O}(r^3)~.
\end{equation}
\textit{$8^{th}$ order:} Here, the form of the  inner solution is
\begin{align}
    f_{b}^{8}(\tilde{r},\theta) =\underbrace{ \left[3\cos(\theta ) +5\cos(3\theta)\right] \frac{A_{8}}{\tilde{r}^{4}}}_{\text{must match } F_{b}^{10}} ~+~ \underbrace{\left[3\cos(\theta ) +5\cos(3\theta)\right]D_1\tilde{r}^{3}}_{\text{matching } F_{b}^{3}} \ , \
    f_{s}^{8}(\theta) = \tilde{f}_{s}^{8}\left[3\cos(\theta ) +5\cos(3\theta)\right] ~.
\end{align}
As in the case of the $5^{th}$ order, these terms do not contribute to the expression of $D_e$. There are no matching term for the outer solution so it simply reads $F_{b}^{8} = 0$. Finally, the inner solution at $9^{th}$ order reads
\begin{align}
    f_{b}^{9}(\tilde{r}, \theta) &= \underbrace{\frac{A_{9}}{\tilde{r}^{2}}\cos(\theta)}_{\text{must match } F_{b}^{9}} +~ \underbrace{B_{9}\tilde{r}\cos(\theta)}_{\text{matching } F_{b}^{6}}~+~ \underbrace{E_{0}(\theta,\phi)\tilde{r}^{7}}_{\text{matching } F_{b}^{0}} ~+ \underbrace{\frac{C_9 E_{0}(\theta,\phi)}{\tilde{r}^{8}} }_{\text{must match } F_{b}^{15}} ~,\\
    f_{s}^{9}(\theta) &= \tilde{f}_{s}^{9}\cos(\theta )~, \ A_{9}=\frac{B_{9}}{2}  \alpha~, \ B_{9} = -\frac{4\pi}{3} A_{6}~,\  \tilde{f}_{s}^{9} =  \frac{3 \xi D_b B_{9} }{2\left(D_s \xi (2\gamma +1)+D_b\right)}~.
\end{align}
Since $E_{0}$ satisfies Eq. (\ref{legendre}), the two last terms in $f_{b}^{9}$ do not contribute in the expression of $D_e$.

We stop here the iteration procedure and we consider the long-time dispersion Eq.~(\ref{deff_obstacles_3D}) which reads:
\begin{align}
                 D_{e} =  P_{b}^{st,0}\left\{\{D_b(1-\varphi) + 2 D_s\xi \varphi   +  D_b \varphi\left[A_{0} + \sum_{n=1}^3 R^{3n} (A_{3n} +B_{3n})  \right] 
        -2D_{s}\varphi  \sum_{n=0}^3 R^{3n}\tilde{f}_{s}^{3n}  \right\}+o(R^{12})~.
\end{align}
Using the above results for $A_n,B_n,\tilde{f}_{s}^{ n}$, this expression can be simplified, leading to a re-summation and to the  result 
\begin{equation}
    D_{e} = D_b P_{b}^{st,0} \ \left\{1- \frac{3\varphi\alpha}{2}\left[1-\frac{\alpha \varphi}{2} +\left(\frac{\alpha \varphi}{2}\right)^{2}- \left(\frac{\alpha \varphi}{2}\right)^{3} \right]\right\} + o(\varphi^4)  = \frac{D_b}{1+ (3\xi-1)\varphi} \ \frac{1-\alpha \varphi}{1 + \frac{\alpha \varphi}{2}} +o(\varphi^4)\label{obstacles_3D}~.
\end{equation}

\subsection{Comparison with existing results }
\subsubsection{Monte Carlo  data of Ref.~\cite{putzel2014nonmonotonic} }

Here, we compare the prediction of Eq.~(15)  with Monte Carlo simulations of Ref.~\cite{putzel2014nonmonotonic}, as they appear in Figure 9 of their Supplementary Information, for periodic arrays of weakly sticky crowders. The result of this comparison is shown in Fig.2, and here we describe briefly how to identify the  parameters of our surface-mediated diffusion model with their short range interaction model. In Ref.~\cite{putzel2014nonmonotonic}, the tracer particle is subject to an exponentially decaying potential $U(D)$, with $D$ the distance to the crowder's wall. The link between the parameter $\delta$ in our description and the potential $U(D)$ can be made by considering the simple one-dimensional situation in which the tracer particle  diffuses between $x=0$ and $x=H$ and is subject to the potential $U(x)$, assumed to vanish for distances larger than a length $\lambda$, with $\lambda \ll H$. In this case, the probability $p_{\Sigma}$ to find the particle at a distance  from the surface less than $\varepsilon$ (with $\lambda \ll \epsilon \ll H$) reads
\begin{equation}
    p_{\Sigma} =\frac{ \int_{0}^{\epsilon} e^{-\beta U(x)} \dd x}{\int_{0}^{H} e^{-\beta U(x)} \dd x  }
    = \frac{ \int_{0}^{\epsilon} [e^{-\beta U(x)}-1] \dd x +\varepsilon}{\int_{0}^{H} [e^{-\beta U(x)}-1] \dd x +H }\
    \simeq \dfrac{\int_{0}^{\infty} \left( e^{-\beta U(x)} -1 \right) \dd x }{\int_{0}^{\infty} \left( e^{-\beta U(x)} -1 \right) \dd x + H}~.
\end{equation}  
In our formalism, we  readily find  $p_{\Sigma}=\delta/(\delta+H)$, so that the expression of the length $\delta$  in terms of $U$ is
\begin{equation}
    \delta = \int_{0}^{\infty} \left( e^{-\beta U(D)} -1 \right) \dd D~. \label{EqDelta}
\end{equation}  
Next, since there is no potential barrier between the surface bound state and the bulk, we are in the fast exchange limit, $k_a,k_d\to\infty$, leading to $\gamma=0$ in Eq.~(15). Finally, Ref.~\cite{putzel2014nonmonotonic} considers a tracer particle of finite radius $r_d$ in presence of spherical crowders of radius $r_c$ and volume fraction $\phi$, this is equivalent to considering a punctual tracer particle in presence of crowders of radius $R=r_c+r_d$, and thus the volume fraction in our description is $\varphi=\phi(1+r_d/r_c)^3$.  
We obtain $\varphi = 0.57$, for this rather high value the obstacles are actually touching each other, but this is physical since the crowders are actually not touching each other.  It is actually very surprising that our formula can correctly predicts $D_e$ for these parameters, as found in  Fig.  ~\ref{fig_deff_obstacles}(left) and Fig. 2 in the main text. 
   
\subsubsection{Comparison with the result of macrotransport theory  }

In Ref.~\cite{edward1995diffusion}, using the formalism of macrotransport theory, a formula was derived for the dispersion of tracers in a periodic array of spheres in a model where tracers can diffuse inside the sphere, and in addition  undergo surface diffusion or diffusion in the bulk. The boundary conditions at the interface in their approach is $p_s(\ve[r]_s,t)=k_{\text{ed}} p_b(\ve[r]_s,t)$ and $p_\text{inside}(\ve[r]_s,t)=K_{\text{ed}}p_b(\ve[r]_s,t)$ where $k_{\text{ed}}$ and $K_{\text{ed}}$ are Henry's laws constants, and $p_\text{inside}(\ve[r],t)$ is the probability density of tracer position inside the spheres. In principle, if one takes $K_{\text{ed}}=0$ to prevent the tracers from entering into the obstacles, the physical situation of Ref.~\cite{edward1995diffusion} is identical to our formalism in the fast exchange limit ($\gamma=0$). The result of Ref.~\cite{edward1995diffusion} for a periodic array of spheres in 3D is given by their equation (4.23)  and reads with our notations:
\begin{equation}
   D_e  = \frac{D_b}{1+ (3\xi -1) \varphi } \left[ 1- \frac{3 \varphi}{2\left(1+ \frac{D_s}{D_b} \xi \right) + \varphi \left(1-2\frac{D_s}{D_b}\xi \right)}\right]+~ o(\varphi^4) = \frac{D_b}{1+ (3\xi-1)\varphi} \ \frac{1- \varphi}{1 + \frac{\alpha \varphi}{2}} +~ o(\varphi^4)\label{edwards}~, 
\end{equation}
which is identified by writing $K_{\text{ed}} = \gamma_{\text{ed}} = 0$, $k_{\text{ed}}\alpha_{\text{ed}} = 3 \xi \varphi$, $\Gamma_{\text{ed}} = D_s\xi /D_b$ (the suffix ``$\text{ed}$'' refers to notations of Ref. \cite{edward1995diffusion}). However, the above equation (\ref{edwards}) disagrees with our Eq. (15), since the numerator contains a term $1- \varphi$ rather  than $1-\alpha\varphi$.    

\begin{figure}[h!]
    \includegraphics[width=8 cm]{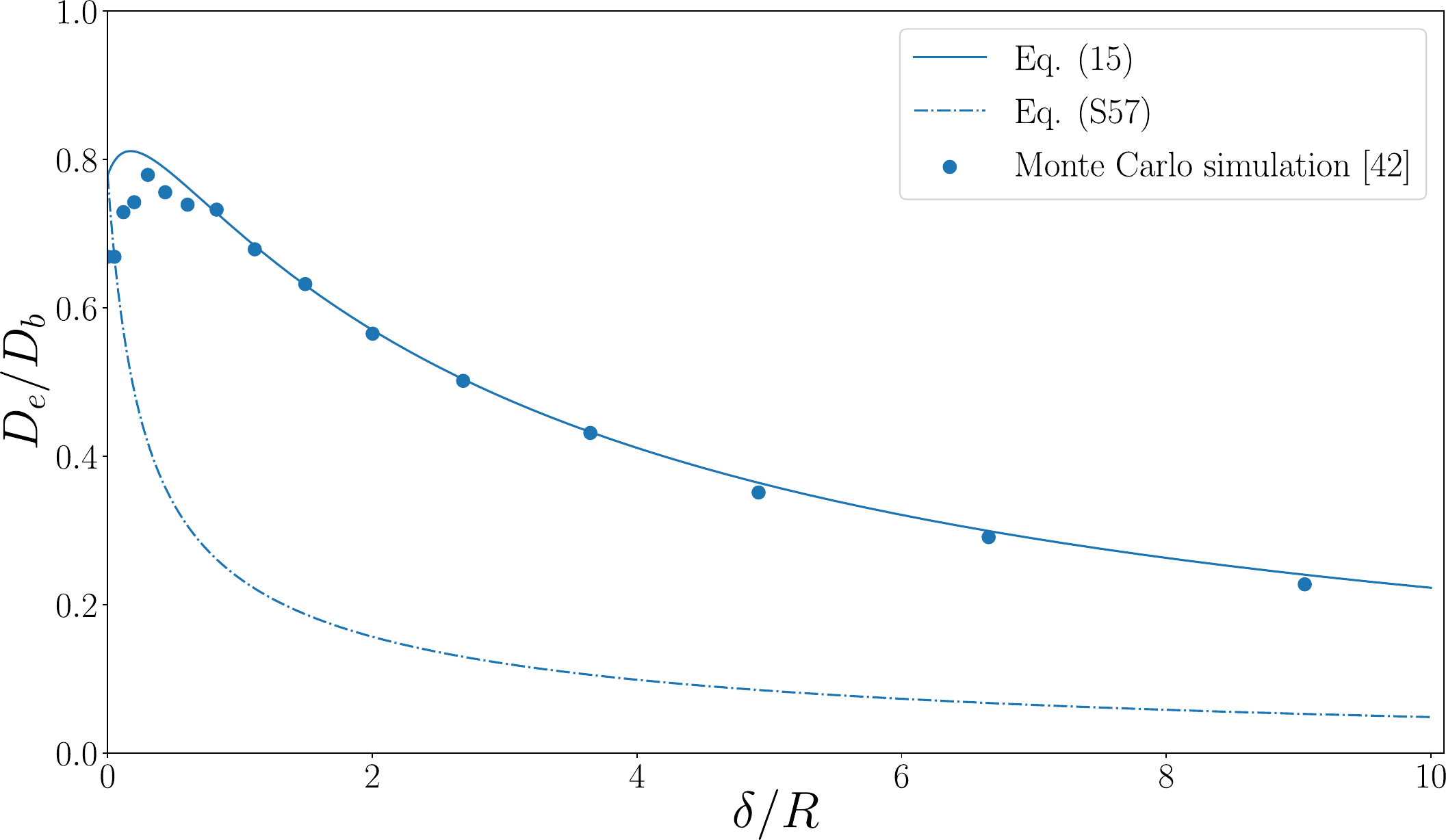}
    \includegraphics[width=8cm]{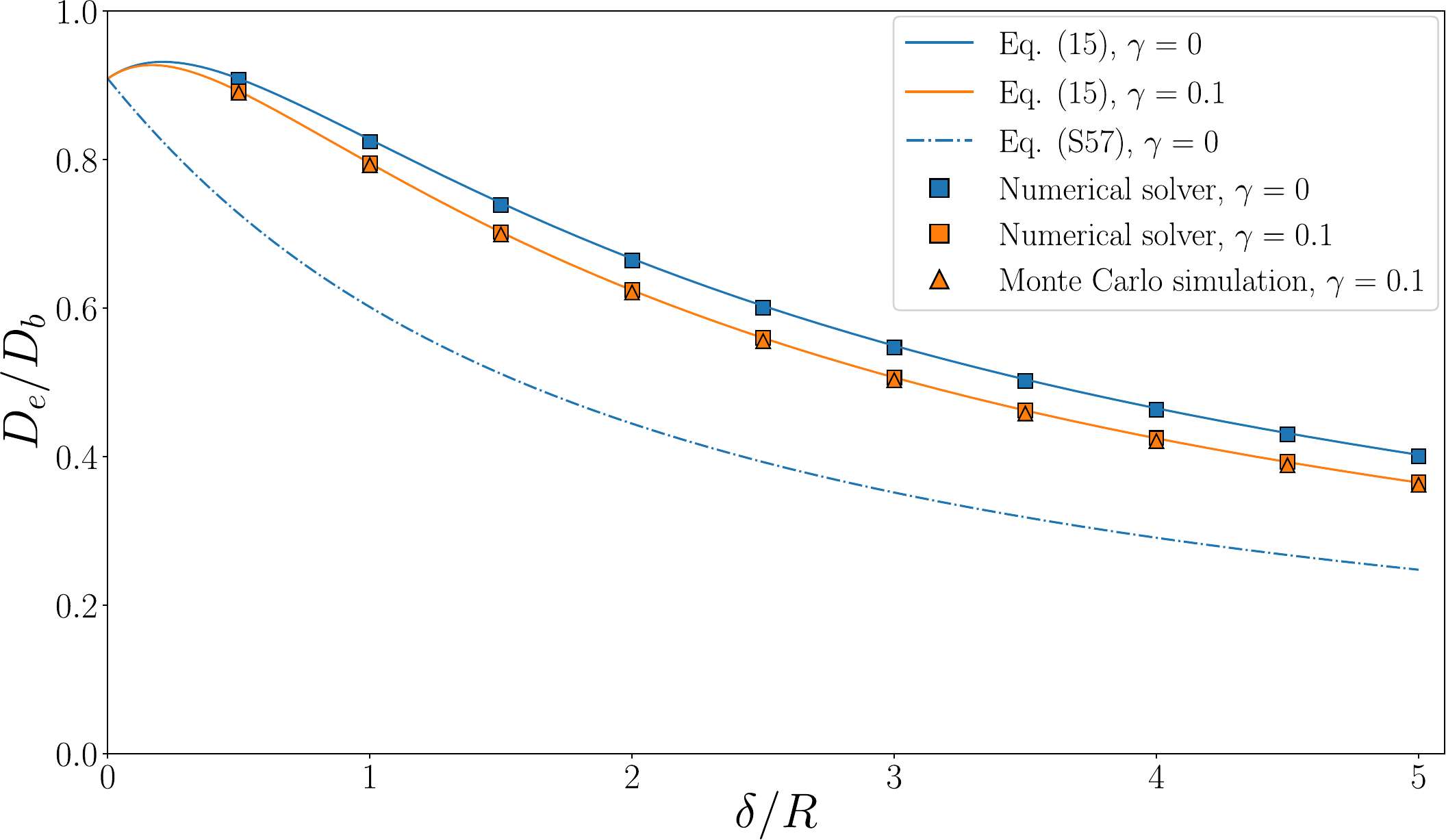}
    \caption{Left:  Comparison between Monte Carlo simulation data for dispersion in 3D arrays of obstacles (circles: data of Fig. 9 in the SI of     Ref.~\cite{putzel2014nonmonotonic}), our theoretical prediction and the result (\ref{edwards}) of Ref.~\cite{edward1995diffusion}. Parameters: $\varphi =0.57$, $D_s=D_b$,  $\gamma = 0$, and $\delta$ is identified using Eq.~(\ref{EqDelta}) for an exponentially decaying potential $U(D)$. Right. Comparison between the data obtained from the numerical resolution of our formalism (squares), stochastic simulations (triangles) and analytical predictions (see legend). Parameters: $\varphi =0.2$, $D_s=D_b$,   $\gamma = 0$ (blue) or  $\gamma = 0.1$ (orange) and a time step $\dd t = 10^{-6} L^2/D_b$.}
    \label{fig_deff_obstacles}
\end{figure}

In Fig.~\ref{fig_deff_obstacles}, we show our prediction [Eq. (15)] along with  the result  (\ref{edwards}), compared with the numerical solution of our formalism. Our analytical formula agrees with the numerical results, while  the formula \ref{edwards} does not and in particular the non-monotonicity of $D_e$ with $\delta$ is absent.  An additional  check of the validity of our formalism is provided by comparison with direct  stochastic numerical simulations of the  process $\ve[r](t)$ corresponding to the Fokker-Planck equations (1) and (2) (described in Section \ref{SectionSimulations}). The values of $D_e$ directly measured from the  stochastic trajectories obtained with this algorithm are shown on Fig.~\ref{fig_deff_obstacles}(right) and confirm the validity of our formalism.

\begin{figure}[h!]
	\includegraphics[width=8cm]{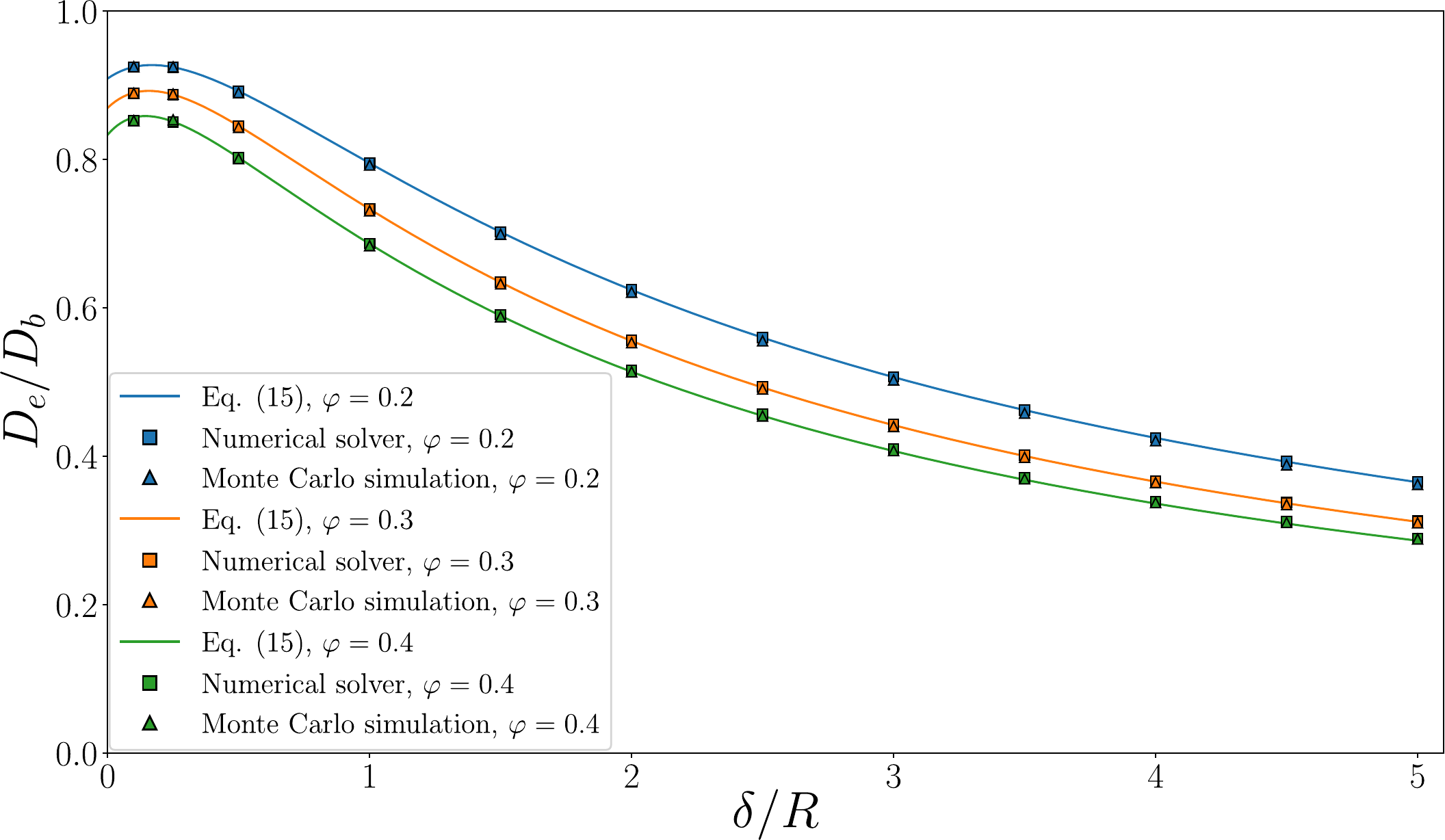}
	\caption{Data obtained for spherical obstacles from numerical integration of our formalism (squares), stochastic simulations (triangles) and analytical predictions for different values of $\varphi$ (see legend). Parameters: $D_s=D_b$, $\gamma = 0.1$ and $\dd t = 10^{-6} L^2/D_b$.}
	\label{fig_deff_varphi}
\end{figure}

 Furthermore, the above results for the diffusivity Eqs.~(\ref{obstacles_2D_final}) and (\ref{obstacles_3D}) are in surprisingly good agreement with simulations even for higher values of $\varphi$, as shown in Fig.~\ref{fig_deff_varphi} in the case of spherical obstacles.

\section{Dispersion in slowly undulated channels ($a\ll L$) - Proof of Eq. (18)}
\label{SlowChannel}
\subsection{The 2D symmetric channel}

In this section, we determine the effective diffusivity $D_e$ in a two-dimensional symmetric channel of height $h(x)$ satisfying $h(x+L) = h(x)$, in the limit of large $L$. The channel wall is parametrized by $x$, we note that $\ve[e]_{s,x}=\ve[e]_x+h'(x)\ve[e]_y$, so that $g_{xx}=1+h'(x)^2$, $\nabla_s^2x=-h'(x)h''(x)/\left[1+h'(x)^2\right]^2$. Using the formulas of Section \ref{MetricSection}, Eq.~(12, 13) read
\begin{equation}\left\{
    \begin{split}
        &  \frac{\partial^{2} f_{b}}{\partial x^{2}}+\frac{\partial^{2} f_{b}}{\partial y^{2}} = 0~, \hspace{1cm}& \left[0<y<h(x)\right] \\                
        &   \frac{D_s}{\sqrt{1+h'(x)^2}}\frac{\partial}{\partial x}\left[ \frac{1}{\sqrt{1+h'(x)^2}}\frac{\partial f_s}{\partial x} \right]-k_d f_s+k_a f_b= - D_s P_s^{st,0} \frac{h'(x)h''(x)}{[1+h'(x)^2]^2}~, \hspace{1cm}& \left[y=h(x)\right] \\
        & \frac{ D_b}{\sqrt{ 1 + h'(x)^2}} \left[\frac{\partial f_{b}}{\partial y} - h'( {x}) \frac{\partial f_{b}}{\partial x} \right]
        = k_d f_{s} - k_a f_{b}  -\frac{D_b P_b^{st,0}h'(x)}{\sqrt{1+ h'(x)^{2}}}~.\hspace{1cm}& \left[y=h(x)\right]
    \end{split}
    \right.
\end{equation}
Furthermore, $f_b$ and $f_s$ are periodic with period $L$, and satisfy $\partial_yf_b\vert_{y=0}=0$ (from the  symmetry $y\to-y$). The stationary probability densities and the effective diffusivity are given by
\begin{align}
&P_b^{st,0}=\frac{1}{V+S\delta}~, \hspace{1cm}  P_s^{st,0} = \delta P_b^{st,0}~,\hspace{1cm}  V=2\int_0^L dx \ h(x)~, \hspace{1cm}S= 2\int_0^L dx \sqrt{1+h'(x)^2}~,\\
 &       D_{e}  = P_{b}^{st,0} V D_{b} + 2D_{s}P_{s}^{st,0} \int_{0}^{L} \frac{dx}{\sqrt{1+h'(x)^{2}}}  
         -D_{b}\int_{0}^{L}dx\int_{-h(x)}^{h(x)}dy\frac{\partial f_{b}}{\partial x}        -2D_{s}\int_{0}^{L}\frac{dx}{\sqrt{1+ h'(x)^{2}}}\frac{\partial f_{s}}{\partial x} \label{De2DChannel}~.
    \end{align}

We now analyze the above equations in the limit $L\to\infty$, when all other parameters are held constant. Here, we write $h(x)=\zeta(x/L)$, where the function $\zeta(\tilde{x})$ does not depend on $L$, and we use the notation $\tilde{x}=x/L$.   The expansion of $f_b$ takes the form:
\begin{align}
f_b(\ve[r])=f_b^0(\tilde{x},y)+L^{-1} \ f_b^1(\tilde{x},y)+L^{-2}\ f_b^2(\tilde{x},y)+...~, \hspace{0.5cm} f_s(x)=f_s^0(\tilde{x})+L^{-1} \ f_s^1(\tilde{x})+L^{-2}\ f_s^2(\tilde{x},y)+...\label{Ansatz2D}
\end{align}
Note that this expansion in powers of $L$ is equivalent to an expansion in terms of the small parameter $\varepsilon=a/L$, with $a$ the minimal channel width. This kind of calculation is very similar to those in lubrication theory in hydrodynamics.
In order to perform the expansion it is useful to consider the equations in terms of the rescaled variable $\tilde{x}=x/L$:
\begin{equation}
\left\{
    \begin{split}
        &  \frac{1}{L^2}\frac{\partial^{2} f_{b}}{\partial \tilde{x}^{2}}+\frac{\partial^{2} f_{b}}{\partial y^{2}} = 0~, & \left[ 0<y<\zeta(\tilde{x})\right] \\
        & \frac{D_s  }{1+  \zeta'(\tilde{x})^2/L^2}\left[\frac{1}{L^2}\frac{\partial^{2} \tilde{f}_{s}}{\partial \tilde{x}^{2}} -\frac{1}{L^4} \frac{\zeta'(\tilde{x})\zeta''(\tilde{x})}{1 +  \zeta'(\tilde{x})^2/L^2}\frac{\partial \tilde{f}_{s}}{\partial \tilde{x}} \right] =
        k_d  f_{s}-k_a f_{b}-   \frac{D_s  P_{s}^{st, 0}\zeta'(\tilde{x})\zeta''(\tilde{x})}{L^3\left(1+\zeta'(\tilde{x})^2/L^2\right)^2}~, &  \left[ y=\zeta(\tilde{x})\right] \\
        & \frac{ D_b }{\sqrt{ 1 + \zeta'(\tilde{x})^2/L^2}}\left[\frac{\partial f_{b}}{\partial y} - \frac{\zeta'(\tilde{x})}{L^2}\frac{\partial f_{b}}{\partial \tilde{x}}\right]\left[\tilde{x},h(\tilde{x})\right]
        = k_d f_{s} -k_a f_{b} 
        - \frac{D_b P_b^{st, 0}\zeta'(\tilde{x})}{L\sqrt{ 1 + \zeta'(\tilde{x})^2/L^2}}~.& \left[ y=\zeta(\tilde{x})\right]
    \end{split}
    \right.
\end{equation}
  \begin{align}
    P_{b}^{st,0} = \frac{1}{C L}\left(1+\mathcal{O}\left( L^{-2}\right)  \right)~, \hspace{1cm} P_{s}^{st,0} = \frac{\delta}{CL}\left(1+\mathcal{O}\left(L^{-2}\right)  \right)~, \hspace{1cm}C=2 \left(\left<\zeta \right> +\delta \right)~.
\end{align}
We insert the ansatz (\ref{Ansatz2D}) into the above equations to find  successive orders $1/L^n$.
\textit{At leading order}, we obtain
\begin{align}
    &\frac{\partial^{2} f_{b}^{0}}{\partial y^{2}} = 0~, \hspace{1cm}   (f_{s}^{0}- \delta f_{b}^{0})_{y=\zeta(\tilde{x})} = 0~, \hspace{1cm}
     \left(\frac{D_b}{k_d}\frac{\partial f_{b}^{0}}{\partial y}   -  f_{s}^{0}+ \delta f_{b}^{0}\right)_{y=\zeta(\tilde{x})}=0~.\label{boundary_0}
\end{align}
Taking into account the condition $\partial_yf_b^0\vert_{y=0}=0$, the above equations indicate that $f_b^0$ and $f_s^0$ do not depend on $y$, and that they are proportional at the surface:
\begin{equation}
        f_{b}^{0}(\tilde{x},y) = f_{b }^{0}(\tilde{x}) \ , \ 
        f_{s}^{0}(\tilde{x}) = \delta f_{b }^{0}(\tilde{x})~.
    \label{kubo_0} 
\end{equation}
Next, at first order in $1/L$, we obtain the same equations as (\ref{kubo_0}), leading to $ f_{b}^{1}(\tilde{x},y) = f_{b}^{1}(\tilde{x}) $ and 
$        f_{s}^{1}(\tilde{x}) = \delta f_{b }^{1}(\tilde{x}).$ At second order, we obtain
\begin{align}
    \frac{\partial^{2} f_{b }^{0} }{\partial \tilde{x}^{2}}+\frac{\partial^{2} f_{b}^{2}}{\partial y^{2}} &= 0~, &[0<y<\zeta(\tilde{x})] \label{bulk2} \\
    D_s   \frac{\partial^{2} f_{s }^{0}}{\partial \tilde{x}^{2}}  - k_d
     f_{s}^{2}+k_a  f_{b}^{2} &= 0~, & [y=\zeta(\tilde{x})] \label{Surf21}\\ D_b \left[\frac{\partial f_{ b}^{2}}{\partial y}-\frac{\zeta'(\tilde{x})^2}{2}\frac{\partial f_{ b }^{0}}{\partial y} - \zeta'(\tilde{x})\frac{\partial f_{b }^{0}}{\partial \tilde{x}}\right]
    &= \left(k_d f_{s}^{2}-k_a f_{b }^{2}\right)  -\frac{D_b }{C} \zeta'(\tilde{x})~. & [y=\zeta(\tilde{x})] \label{Surf22}
\end{align}
We see that Eq.~(\ref{bulk2}) simplifies to
\begin{equation}
    \frac{\partial f_{b}^{2}}{\partial y}(\tilde{x},y) = -{f_{b }^{0}}''(\tilde{x})y~, \label{integrate_fomega2}
\end{equation}
where we have used the symmetry $y\to-y$. Next, combining Eqs. (\ref{Surf21}), (\ref{Surf22}) and (\ref{integrate_fomega2}) and using the result at order $0$ [Eq. (\ref{kubo_0})], we obtain 
\begin{equation}
    {f_{b}^{0}}''(\tilde{x})\left(D_b \zeta(\tilde{x}) + D_s\delta  \right)+ {f_{ b }^{0}}'(\tilde{x})D_b \zeta'(\tilde{x}) = \frac{D_b \zeta'(\tilde{x})}{C}~.
\end{equation}
The solution of this first-order differential equation that satisfies the periodic boundary conditions is
\begin{equation}
    {f_{b}^{0}}'(\tilde{x})= \frac{1}{C\left(D_b \zeta(\tilde{x})+D_s \delta\right)}\left[ D_b \zeta(\tilde{x}) - \frac{\left<  D_b \zeta/\left(D_b \zeta +  D_s \delta \right) \right>}{\left<\left(D_b \zeta+ D_s \delta \right)^{-1}\right>}\right]~.
\end{equation}
Inserting this result into Eq.~(\ref{De2DChannel}), we find (after a few lines of algebra)
 \begin{equation}
         D_e = \frac{1}{\left<h + \delta \right>\left<\left(D_{b} h+ D_s \delta \right)^{-1}\right>} + \mathcal{O} (L^{-2}) \label{D_LJ_2D}~.
\end{equation}
  
\subsection{The 3D axisymmetric channel}
 
We now consider an axisymmetric channel oriented along the $x$ direction of  local radius $h(x)$. We denote by $r$ the distance to the central axis and $\theta$ the angle around this axis. due to rotational invariance, $f_b(r,x,\theta)$ and $f_s(x,\theta)$ do not depend on $\theta$. 
We parametrize the surface with $x$ and $\theta$. The metric is identified by calculating
 $\textbf{e}_{s, x} = \partial_x \textbf{r}_{s} = \textbf{e}_{x} +h'(x )\textbf{e}_{r}$ and $\textbf{e}_{s, \theta} = h(x)\textbf{e}_{\theta}$ with $ \textbf{e}_{r} $ the unit vector perpendicular to the channel axis, and  $ \textbf{e}_{\theta} =  \textbf{e}_{x} \times \textbf{e}_{r}$. We thus obtain $g_{xx}=1+h'(x)^{2}$, $g_{x,\theta}=0$ and $g_{\theta\theta}= h(x)^{2}$. Using  Section \ref{MetricSection}, Eqs. (12, 13) become 
\begin{equation}\left\{
    \begin{split}
        & \frac{1}{r}\frac{\partial}{\partial r}\left( r\frac{\partial f_{b}}{\partial r} \right) +  \frac{\partial^{2} f_{b}}{\partial x^{2} } = 0~,\\
        & \frac{D_s}{1+h'(x)^2} \left\{ \frac{\partial^{2} f_{s}}{\partial x^{2}} +   \left[\frac{h'(x)}{h(x)}- \frac{h'(x)h''(x)}{1+h'(x)^2} \right]\frac{\partial f_{s}}{\partial x} \right\}  =
        k_d f_{s}- k_a f_{b} 
        + \frac{D_s  P_{s}^{st,0} }{ 1 + h'(x)^2 } \left\{\frac{h'(x)}{ h(x) }   -\frac{h'(x)h''(x)}{ 1 + h'(x)^2}\right\}~,\\
        & \frac{D_b}{\sqrt{ 1+ h'(x)^2}}\left[ \frac{\partial f_{b}}{\partial r} - h'(x) \frac{\partial f_{b}}{\partial x}\right]\left[x,h(x)\right] =k_d f_{s}- k_a f_{b}
        - \frac{ D_b P_{b}^{st,0} h'(x)}{ \sqrt{ 1+ h'(x)^2}}~,
    \end{split}\right.
\end{equation}
and $D_e$ is given by
\begin{equation}
    D_{e}  = P_{b}^{st,0} V D_{b} + 2\pi D_{s}P_{s}^{st,0} \int_{0}^{L} \frac{dx h(x)}{\sqrt{1+h'(x)^{2}}}  
    - 2\pi D_{b}\int_{0}^{L}dx\int_{0}^{h(x)}dy y\frac{\partial f_{b}}{\partial x}        -2\pi D_{s}\int_{0}^{L}\frac{dx h(x)}{\sqrt{1+ h'(x)^{2}}}\frac{\partial f_{s}}{\partial x} \label{De3DChannel}~.
\end{equation}
As in the 2D case, we set  $h(x) = \zeta(x/L)$ and we use the notation $\tilde{x}=x/L$, which gives
\begin{equation}\left\{
    \begin{split}
    &\frac{1}{r}\frac{\partial}{\partial r}\left( r\frac{\partial f_{b}}{\partial r} \right) + \frac{1}{L^2} \frac{\partial^{2} f_{b}}{\partial \tilde{x}^{2} } = 0~,\\
        &\frac{D_s}{L^2 \zeta'(\tilde{x})^2} \left\{ \frac{\partial^{2} f_{s}}{\partial \tilde{x}^{2}} +   \left[\frac{\zeta'(\tilde{x})}{\zeta(\tilde{x})}- \frac{\zeta'(\tilde{x})\zeta''(\tilde{x})}{L^2 +\zeta'(\tilde{x})^2 } \right]\frac{\partial f_{s}}{\partial \tilde{x}} \right\}   =
        k_d f_{s}- k_a f_{b} 
        +\frac{ L D_s  P_{s}^{st,0} } { L^2 + \zeta'(\tilde{x})^2}\left\{\frac{\zeta'(\tilde{x})}{\zeta(\tilde{x})  }   -\frac{\zeta'(\tilde{x})\zeta''(\tilde{x})}{\left[L^2 + \zeta'(\tilde{x})^2\right]}\right\}~,\\
    &\frac{D_b}{\sqrt{ 1+ \zeta'(\tilde{x})^2/L^2}}\left[ \frac{\partial f_{b}}{\partial r} - \frac{\zeta'(\tilde{x})}{L^2} \frac{\partial f_{b}}{\partial \tilde{x}}\right]_{\left[\tilde{x},\zeta(\tilde{x})\right]} =k_d f_{s}- k_a f_{b}
    - \frac{ D_b P_{b}^{st,0} \zeta'(\tilde{x})}{ L \sqrt{ 1+ \zeta'(\tilde{x})^2/L^2}}~,
\end{split}\right.
\end{equation}
with 
\begin{align}
    P_{b}^{st,0}=\frac{1}{C'L} \left[ 1 + \mathcal{O} \left( L^{-2} \right)\right] ~, \
    P_{s}^{st,0}=\frac{\delta}{C'L} \left[ 1 + \mathcal{O}  \left( L^{-2} \right)\right]~,\ C'=\pi \left< h^{2}\right>+2\pi \delta\left< h\right>~.
\end{align}
In the limit $L\to\infty$ the solutions have a series expansion [see Eq. (17)] and  successive orders are found by applying the same method as for the   2D channel. As in the 2D case,  at leading and first order, we find
\begin{equation}
        f_{b }^{0}(\tilde{x},r) = f_{b }^{0}(\tilde{x})~, \ 
        f_{s}^{0}(\tilde{x}) = \delta f_{b }^{0}(\tilde{x})~,        \
         f_{b }^{1}(\tilde{x},r) = f_{b }^{1}(\tilde{x})~, \ 
        f_{s}^{1}(\tilde{x}) = \delta f_{b }^{1}(\tilde{x}) ~.
\end{equation}
The equations at $2^{nd}$ order are
\begin{align}
    &\frac{1}{r}\frac{\partial}{\partial r}\left( r\frac{\partial f_{b }^{2}}{\partial r} \right) + \frac{\partial^{2} f_{b }^{0}}{\partial \tilde{x}^{2}} = 0~, & \left[ 0 <r< \zeta(\tilde{x})\right] \label{bulk_2_3D}\\&
    D_s \left[\frac{\partial^{2} f_{s}^{0}}{\partial \tilde{x}^{2}} + \frac{\zeta'(\tilde{x})}{\zeta(\tilde{x})} \frac{\partial f_{s}^{0}}{\partial \tilde{x}}\right]
    = k_d f_{s}^{2} - k_a f_{b }^{2}  + \frac{D_s \delta \zeta'(\tilde{x})}{C' \zeta(\tilde{x})}~,\label{surface_2_3D}&\left[ r=\zeta(\tilde{x})\right]\\&
    D_b \left[\frac{\partial f_{b }^{2}}{\partial r} - \frac{\zeta'(\tilde{x})^2}{2 }\frac {\partial f_{b }^{0}}{\partial r} -\zeta'(\tilde{x}) \frac{\partial f_{b}^{0} }{\partial \tilde{x}}\right] =  k_d f_{s}^{2} -k_a f_{b}^{2} 
    -\frac{D_b \zeta'(\tilde{x})}{C'}~. &\left[ r=\zeta(\tilde{x})\right]\label{boundary_2_3D}
\end{align}
The integration of Eq.~(\ref{bulk_2_3D}) leads to $\partial_r f_{b}^{2} = -\frac{r{f_{ b}^{0}}''(\tilde{x})}{2}$ which can be combined with  Eqs.~(\ref{surface_2_3D}, \ref{boundary_2_3D}), leading to
 \begin{equation}
    {f_{b}^{0}}''(\tilde{x}) \left[ \frac{D_b \zeta^2(\tilde{x})}{2}+D_s \delta \zeta(\tilde{x}) \right]+
    {f_{b}^{0}}'(\tilde{x})\left[ D_b \zeta(x) \zeta'(\tilde{x})+ D_s \delta \zeta'(\tilde{x}) \right] = \frac{1}{C'} \left[ D_b \zeta(\tilde{x})\zeta'(\tilde{x}) +D_s \delta \zeta'(\tilde{x}) \right]~.
\end{equation}
Solving this differential equation, we obtain:
\begin{equation}
    {f_{b}^{0}}'(\tilde{x}) = \frac{1}{C'}\left[ 1- \frac{1}{\zeta(\tilde{x})\left(D_b \zeta(\tilde{x})+2 D_s \delta \right)\left< h^{-1}\left(D_b h+2 D_s \delta \right)^{-1} \right>}\right]~.
\end{equation}
Inserting this result into Eq.~(\ref{De3DChannel}) yields the   $D_e$ at leading order in $1/L$:
 \begin{equation}
        D_{e} = \frac{1}{\left<h\left( h + 2\delta \right) \right>\left<h^{-1}\left(D_bh + 2 D_s \delta \right)^{-1}\right>} + \mathcal{O}\left(L^{-2}\right)~.
    \label{D_LJ_3D}
\end{equation}
 
\subsection{Upper bounds on critical surface diffusion $D_s^*$} 

Here we show that 
\begin{equation}
    D_s^* ~\leq~ D_s^0 ~\leq ~D_b~, \label{inequality_D_s}
\end{equation}
where we recall that $D_s^*$ is the value of $D_s$ for which $\partial D_e/\partial\delta=0$, and  $D_s^0$ is the value of $D_s$ for which $D_e(\delta=0)=D_e(\delta=\infty)$. Using Eq.~(18) it is straight forward to show that
\begin{align}
   & D_s^* =D_b  \frac{\left< h^{d-2}\right>\left< h^{1-d}\right>}{\left< h^{d-1}\right>\left< h^{-d}\right>}~,&\ D_s^0  = D_b\frac{\left< h^{d-2}\right>\left< h^{2-d}\right>}{\left< h^{d-1}\right>\left< h^{1-d}\right>}~.
\end{align}
Here, we apply the Cauchy-Schwarz inequality which states that, for all integrable functions $ f, g$:
\begin{align}
    \left< fg \right>^2 \leq  \left< f^2 \right> \left< g^2 \right>~, \label{cauchy_schwartz}
\end{align}
The choice $f=h^{1-d/2}$ and $g=h^{-d/2}$ leads to the inequality 
$     \left< h^{1-d}\right>^2 \leq \left< h^{-d}\right>\left< h^{2-d}\right>,$
which leads to   $D_s^* \leq D_s ^0$. Now, using again Eq.~(\ref{cauchy_schwartz}), we find 
  $   \left< h^{d-2}\right>^2 \leq \left< h^{d-1}\right>\left< h^{d-3}\right> $ and 
$     \left< h^{2-d}\right>^2 \leq \left< h^{1-d}\right>\left< h^{3-d}\right>$, 
 leading to
\begin{align}
    \frac{\left< h^{d-2}\right>\left< h^{2-d}\right>}{\left< h^{d-1}\right>\left< h^{1-d}\right>} \leq \frac{\left< h^{d-3}\right>\left< h^{3-d}\right>}{\left< h^{d-2}\right>\left< h^{2-d}\right>} ~.
    \label{induction}
\end{align}
 For $d=2$, the inequality $ 1 \leq \left< h\right> \left< h^{-1}\right>$  directly yields $D_s^0 \leq D_b$. For higher dimensions, we find by induction
\begin{align}
\frac{D_s^0}{D_b}=    \frac{\left< h^{d-2}\right>\left< h^{2-d}\right>}{\left< h^{d-1}\right>\left< h^{1-d}\right>} \leq \frac{\left< h^{d-3}\right>\left< h^{3-d}\right>}{\left< h^{d-2}\right>\left< h^{2-d}\right>} \leq ... \leq \frac{1}{\left< h\right>\left< h^{-1}\right>} \leq 1~,
\end{align} so we recover Eq. (\ref{inequality_D_s}).

\subsection{Physical derivation for dispersion in slowly varying channels}
Here, we see how the above results can be obtained via a physical argument similar to that used for the original derivation of the Fick-Jacobs approximation. For convenience, we consider a channel with a finite size surface region of thickness $\ell$ where there is a uniform potential   $v_s$. In terms of the coordinates perpendicular to the channel axis, the overall  potential $V({\bf r}_\perp)$ is taken to be zero outside the surface region. As the surface varies slowly along the channel, we assume that the position of the particle in the height direction (the plane in the channel at longitudinal  position $x$ denoted $C(x)$) equilibrates giving an $x$ dependent partition function
\begin{equation}
    Z(x) = \int_{C(x)} d{\bf r}_{\perp}e^{-\beta V({\bf r}_{\perp})}= s_{d-1}h(x)^{d-1}+ a_{d-1}h(x)^{d-2}\ell\exp(-\beta v_s)~,
\end{equation}
where $s_{d}$ denotes the volume of a unit $d$ dimensional sphere and $a_d$ its area.
To recover the surface mediated diffusion model considered here one   takes the limit $\ell\to 0$ with
$\ell\exp(-\beta v_s)=\delta$ [see Eq.~(\ref{EqDelta})], and we note that  $a_d=d s_d$. The free energy $\mathcal{F}$ as a function of $x$ is thus given by
\begin{equation}
    \mathcal{F}(x) = -\frac{1}{\beta}\left[ \ln\left(h(x)^{d-1}+ (d-1)\delta h(x)^{d-2}\right)+\ln(s_{d-1})\right]~.
\end{equation}
Also,  we can introduce the average diffusion constant at position $x$ given by
\begin{equation}
    D_a(x) =
    \frac{D_s (d-1)\delta h(x)^{d-2}+ D_b h(x)^{d-1}}{(d-1)\delta h(x)^{d-2}+ h(x)^{d-1}}~.
\end{equation}
The effective diffusion constant in one dimension in a periodic potential $\mathcal{F}(x)$ with periodic diffusion constant $D_a(x)$ is given by the Lifson-Jackson formula \cite{lifson1962self}
\begin{equation}
    D_e=\frac{1}{\langle \exp(-\beta\mathcal{F})\rangle \langle \exp(\beta\mathcal{F})D^{-1}_a\rangle}
    = \frac{1}{\langle h^{d-1}+ (d-1)\delta h^{d-2}\rangle \langle \left(D_b h^{d-1}+D_s (d-1)\delta h^{d-2} \right)^{-1} \rangle}~,
\end{equation}
which is precisely Eq. (18) of our Letter.

\section{Dispersion in highly undulated channels ($a\gg L$) - Proof of Eq. (20)}
\label{FastChannel}

\begin{figure}[h!]
	\includegraphics[width=4cm]{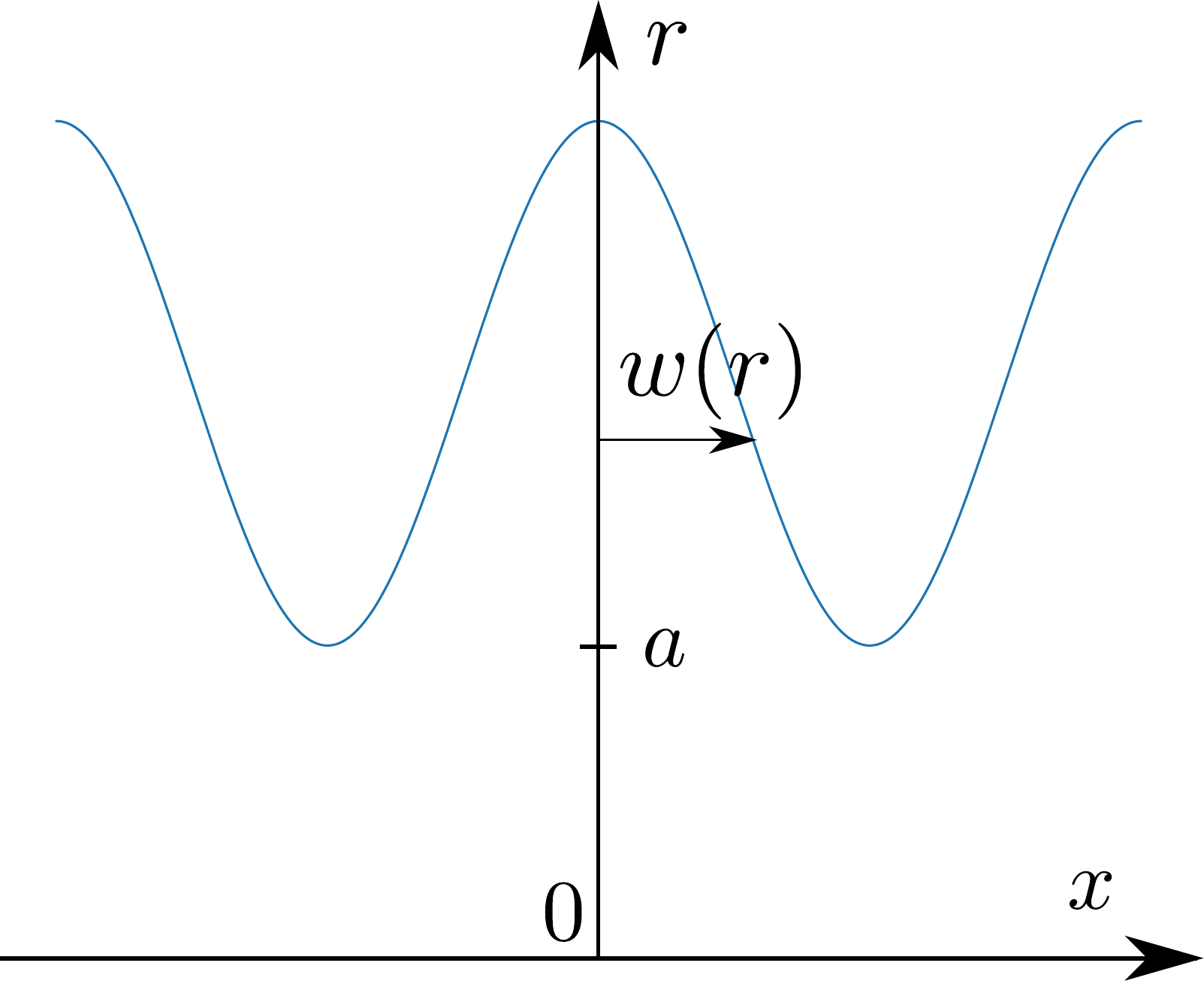}
	\caption{Schematic drawing of a channel where the half-width $w$ is shown.}
	\label{fig_highly_corr_channel}
\end{figure}

This last section concerns channels  in the large corrugation limit.  Let us call $r$ the distance to the central axis, and $w(r)$ the ``half-width'' of the corrugations of the channel, see  Fig.~\ref{fig_highly_corr_channel}. For simplicity, we assume that the corrugation is symmetric about  $x = 0$. This assumption can in fact be avoided by slightly adapting the arguments presented below and does not change the final result.
In 2D, the surface is parametrized with the variable $r$. The metric is determined by calculating $\ve[e]_{s,r} =  \ve[e]_{r} + w'(r) \ve[e]_{x} $ which leads to $g_{rr} = 1+ w'^2(r)$. For an axisymmetric 3D channel, we parametrize the surface with  $r$ and $\theta$, the components of the metric read $g_{rr} = 1+ w'^2(r)$, $g_{r\theta} = 0$ and $g_{\theta \theta} = r^2$. Here, we consider the large corrugation limit $a\to\infty$, by assuming that $w(r) = \tilde{w}( r/a)$, where $\tilde{w}$ does not depend on $a$. In terms of the rescaled variable $\tilde{r}=r/a$, equations (12, 13)  read (for $d=2$ or $d=3$):
\begin{equation}\left\{
    \begin{split}
        &\text{(bulk) }\frac{\partial^2 \tilde{f}_b}{\partial x^2 }+ \frac{1}{a^2 \tilde{r}^{d-2}}\frac{\partial}{\partial \tilde{r}}\left(\tilde{r}^{d-2}\frac{\partial \tilde{f}_b}{\partial \tilde{r}}\right) = 0~, \\
        &\text{(surface) } \frac{D_s }{a^2 \left(1+ \frac{\tilde{w}'^2(\tilde{r})}{a^2}\right)} \left\{ \frac{\partial^2 \tilde{f}_s}{\partial \tilde{r}^2} - \delta \tilde{w}''(\tilde{r})  +\left[ \frac{d-2}{\tilde{r}} - \frac{\tilde{w}'(\tilde{r})\tilde{w}''(\tilde{r})}{a^2\left( 1+ \frac{\tilde{w}'^2(\tilde{r})}{a^2}\right)} \right]\left[\frac{\partial \tilde{f}_s}{\partial \tilde{r}} -\delta \tilde{w}'(\tilde{r}) \right] \right\}  = k_d \tilde{f}_s- k_a \tilde{f}_b ~,\\
        &\text{(b. c.) }\frac{D_b}{\sqrt{1+\frac{\tilde{w}'^2(\tilde{r})}{a^2}}}\left[- \frac{\tilde{w}'(\tilde{r})}{a^2}\frac{\partial \tilde{f}_b}{\partial \tilde{r}} +\frac{\partial \tilde{f}_b}{\partial x} \right] = k_d\tilde{f}_s - k_a \tilde{f}_b+ \frac{D_b}{\sqrt{1+ \frac{\tilde{w}'^2(\tilde{r})}{a^2}}}~,
    \end{split}
    \right.\label{kubo_wide_limit}
\end{equation}
where we used  the notation $\tilde{f}_b = f_b / P_{b, 0}^{st}$ and $\tilde{f}_s = f_s / P_{b, 0}^{st}$. At leading order in the limit $a\to\infty$, we find
  \begin{equation}
     \begin{split}
         \frac{\partial^2 \tilde{f}_b^0}{\partial x^2 } = 0~, \hspace{1cm}
        \left(k_d\tilde{f}_s^0 -k_a \tilde{f}_b^0 \right) \left[\tilde{w}(\tilde{r})~, \tilde{r} \right]= 0~, \hspace{1cm}
         D_b \frac{\partial \tilde{f}_b^0}{\partial x } \left[\tilde{w}(\tilde{r}), \tilde{r} \right] =  k_d \tilde{f}_s^0(\tilde{r} )-k_a\tilde{f}_b^0\left[\tilde{w}(\tilde{r}), \tilde{r} \right]  + D_b~.
     \end{split}\label{kubo_wide_limit0}
 \end{equation}
Solving these equations leads to:
\begin{equation}
    f_{b}^{0}(x,\tilde{r}) = 
    \begin{cases}
        A^{C}(\tilde{r}) &\text{    for } |\tilde{r}|<1, \\
        A^{P}(\tilde{r}) +x &\text{   for } |\tilde{r}|>1,
    \end{cases}
\end{equation}
where the functions $A^C$ and $A^P$  depend only on $\tilde{r}$. The first line of the above expression is the solution in the central region which respects periodic boundary conditions. The second one corresponds to the peripheral region and satisfies the boundary condition given in (\ref{kubo_wide_limit0}). The latter gives a contribution in the expression of the long-time dispersion at leading order in $1/a$. Inserting the solution for $f$ into  the general expression of $D_e$ yields
\begin{equation}
D_e \underset{a\rightarrow+\infty}{=} D_b P_{b, 0}^{st} \left( V - V_p\right),
\end{equation}
with $V_p$ the volume of the peripheral region. Using $P_{b, 0}^{st} = \left[V + \delta S\right]^{-1}$, we thus recover Eq.~(20).

%\bibliography{biblio_heterogeneous_media}
% \bibliography{Biblio_V12}
% SVP NE PAS SUPPRIMER CETTE LIGNE, MAIS LA METTRE EN COMMENTAIRE SI BESOIN
%\bibliography{/Users/thomasguerin/Documents/Recherche/Biblio/Biblio_Guerin}

\begin{thebibliography}{10}
\expandafter\ifx\csname url\endcsname\relax
  \def\url#1{\texttt{#1}}\fi
\expandafter\ifx\csname urlprefix\endcsname\relax\def\urlprefix{URL }\fi
\providecommand{\bibinfo}[2]{#2}
\providecommand{\eprint}[2][]{\url{#2}}

\bibitem{marbach2018transport}
\bibinfo{author}{Marbach, S.}, \bibinfo{author}{Dean, D.~S.} \&
  \bibinfo{author}{Bocquet, L.}
\newblock \bibinfo{title}{Transport and dispersion across wiggling nanopores}.
\newblock \emph{\bibinfo{journal}{Nat. Phys.}} \textbf{\bibinfo{volume}{14}},
  \bibinfo{pages}{1108} (\bibinfo{year}{2018}).

\bibitem{aminian2016boundaries}
\bibinfo{author}{Aminian, M.}, \bibinfo{author}{Bernardi, F.},
  \bibinfo{author}{Camassa, R.}, \bibinfo{author}{Harris, D.~M.} \&
  \bibinfo{author}{McLaughlin, R.~M.}
\newblock \bibinfo{title}{How boundaries shape chemical delivery in
  microfluidics}.
\newblock \emph{\bibinfo{journal}{Science}} \bibinfo{pages}{0532}
  (\bibinfo{year}{2016}).

\bibitem{kim2019tuning}
\bibinfo{author}{Kim, W.~K.}, \bibinfo{author}{Kandu{\v{c}}, M.},
  \bibinfo{author}{Roa, R.} \& \bibinfo{author}{Dzubiella, J.}
\newblock \bibinfo{title}{Tuning the permeability of dense membranes by shaping
  nanoscale potentials}.
\newblock \emph{\bibinfo{journal}{Phys. Rev. Lett.}}
  \textbf{\bibinfo{volume}{122}}, \bibinfo{pages}{108001}
  (\bibinfo{year}{2019}).



\bibitem{brenner2013macrotransport}
\bibinfo{author}{Brenner, H.} \& \bibinfo{author}{Edwards, D.~A.}
\newblock \emph{\bibinfo{title}{Macrotransport processes}}
  (\bibinfo{publisher}{Butterworth-Heinemann, Boston}, \bibinfo{year}{1993}).

\bibitem{leBorgne2013stretching}
\bibinfo{author}{Le Borgne, T.}, \bibinfo{author}{Dentz, M.} \&
  \bibinfo{author}{Villermaux, E.}
\newblock \bibinfo{title}{Stretching, coalescence, and mixing in porous media}.
\newblock \emph{\bibinfo{journal}{Phys. Rev. Lett.}}
  \textbf{\bibinfo{volume}{110}}, \bibinfo{pages}{204501}
  (\bibinfo{year}{2013}).

\bibitem{dentz2011mixing}
\bibinfo{author}{Dentz, M.}, \bibinfo{author}{Le~Borgne, T.},
  \bibinfo{author}{Englert, A.} \& \bibinfo{author}{Bijeljic, B.}
\newblock \bibinfo{title}{Mixing, spreading and reaction in heterogeneous
  media: A brief review}.
\newblock \emph{\bibinfo{journal}{J. Contam. Hydrol.}}
  \textbf{\bibinfo{volume}{120}}, \bibinfo{pages}{1--17}
  (\bibinfo{year}{2011}).

\bibitem{barros2012flow}
\bibinfo{author}{Barros, F.~P.}, \bibinfo{author}{Dentz, M.},
  \bibinfo{author}{Koch, J.} \& \bibinfo{author}{Nowak, W.}
\newblock \bibinfo{title}{Flow topology and scalar mixing in spatially
  heterogeneous flow fields}.
\newblock \emph{\bibinfo{journal}{Geophys. Res. Lett.}}
  \textbf{\bibinfo{volume}{39}} (\bibinfo{year}{2012}).

\bibitem{bernate2012stochastic}
\bibinfo{author}{Bernate, J.~A.} \& \bibinfo{author}{Drazer, G.}
\newblock \bibinfo{title}{Stochastic and deterministic vector chromatography of
  suspended particles in one-dimensional periodic potentials}.
\newblock \emph{\bibinfo{journal}{Phys. Rev. Lett.}}
  \textbf{\bibinfo{volume}{108}}, \bibinfo{pages}{214501}
  (\bibinfo{year}{2012}).

\bibitem{Aminian2015}
\bibinfo{author}{Aminian, M.}, \bibinfo{author}{Bernardi, F.},
  \bibinfo{author}{Camassa, R.} \& \bibinfo{author}{McLaughlin, R.~M.}
\newblock \bibinfo{title}{Squaring the circle: Geometric skewness and symmetry
  breaking for passive scalar transport in ducts and pipes}.
\newblock \emph{\bibinfo{journal}{Phys. Rev. Lett.}}
  \textbf{\bibinfo{volume}{115}}, \bibinfo{pages}{154503}
  (\bibinfo{year}{2015}).

\bibitem{taylor1953dispersion}
\bibinfo{author}{Taylor, G.}
\newblock \bibinfo{title}{Dispersion of soluble matter in solvent flowing
  slowly through a tube}.
\newblock \emph{\bibinfo{journal}{Proc. R. Soc. Lon. A}}
  \textbf{\bibinfo{volume}{219}}, \bibinfo{pages}{186--203}
  (\bibinfo{year}{1953}).

\bibitem{dean2007effective}
\bibinfo{author}{Dean, D.~S.}, \bibinfo{author}{Drummond, I.} \&
  \bibinfo{author}{Horgan, R.}
\newblock \bibinfo{title}{Effective transport properties for diffusion in
  random media}.
\newblock \emph{\bibinfo{journal}{J. Stat. Mech.: Theor. Exp.}}
  \textbf{\bibinfo{volume}{2007}}, \bibinfo{pages}{P07013}
  (\bibinfo{year}{2007}).

\bibitem{bhattacharjee2019bacterial}
\bibinfo{author}{Bhattacharjee, T.} \& \bibinfo{author}{Datta, S.~S.}
\newblock \bibinfo{title}{Bacterial hopping and trapping in porous media}.
\newblock \emph{\bibinfo{journal}{Nature communications}}
  \textbf{\bibinfo{volume}{10}}, \bibinfo{pages}{1--9} (\bibinfo{year}{2019}).

\bibitem{carusela2021computational}
\bibinfo{author}{Carusela, M.~F.} \& \bibinfo{author}{Miguel~Rubi, J.}
\newblock \bibinfo{title}{Computational model for membrane transporters.
  potential implications for cancer}.
\newblock \emph{\bibinfo{journal}{Frontiers in Cell and Developmental Biology}}
  \textbf{\bibinfo{volume}{9}}, \bibinfo{pages}{333} (\bibinfo{year}{2021}).

\bibitem{malgaretti2013entropic}
\bibinfo{author}{Malgaretti, P.}, \bibinfo{author}{Pagonabarraga, I.} \&
  \bibinfo{author}{Rubi, M.}
\newblock \bibinfo{title}{Entropic transport in confined media: a challenge for
  computational studies in biological and soft-matter systems}.
\newblock \emph{\bibinfo{journal}{Frontiers in Physics}}
  \textbf{\bibinfo{volume}{1}}, \bibinfo{pages}{21} (\bibinfo{year}{2013}).

\bibitem{burada2009diffusion}
\bibinfo{author}{Burada, P.~S.}, \bibinfo{author}{H{\"a}nggi, P.},
  \bibinfo{author}{Marchesoni, F.}, \bibinfo{author}{Schmid, G.} \&
  \bibinfo{author}{Talkner, P.}
\newblock \bibinfo{title}{Diffusion in confined geometries}.
\newblock \emph{\bibinfo{journal}{ChemPhysChem}} \textbf{\bibinfo{volume}{10}},
  \bibinfo{pages}{45--54} (\bibinfo{year}{2009}).

\bibitem{jac1967}
\bibinfo{author}{Jacobs, M.}
\newblock \emph{\bibinfo{title}{Diffusion processes}}
  (\bibinfo{publisher}{Springer, New-York}, \bibinfo{year}{1967}).

\bibitem{reguera2006entropic}
\bibinfo{author}{Reguera, D.},
\bibinfo{author}{Schmid, G.},
\bibinfo{author}{Burada, P. S.},  
\bibinfo{author}{Rubi, J. M.},  
\bibinfo{author}{Reimann, P.},  
\bibinfo{author}{H{\"a}nggi, P.},  
\newblock \bibinfo{title}{Entropic transport: Kinetics, scaling, and control  mechanisms}.
\newblock \emph{\bibinfo{journal}{Phys. Rev. Lett.}}
  \textbf{\bibinfo{volume}{96}}, \bibinfo{pages}{130603}
  (\bibinfo{year}{2006}).

\bibitem{rubi2019entropic}
\bibinfo{author}{Rubi, J.~M.}
\newblock \bibinfo{title}{Entropic diffusion in confined soft-matter and
  biological systems}.
\newblock \emph{\bibinfo{journal}{Europhys. Lett.}}
  \textbf{\bibinfo{volume}{127}}, \bibinfo{pages}{10001}
  (\bibinfo{year}{2019}).

\bibitem{yang2017hydrodynamic}
\bibinfo{author}{Yang, X.} 
\bibinfo{author}{Liu, C.} 
\bibinfo{author}{Li, Y.} 
\bibinfo{author}{Marchesoni, F.} 
\bibinfo{author}{Hänggi, P.} 
\bibinfo{author}{Zhang, H. P.} 
\newblock \bibinfo{title}{Hydrodynamic and entropic effects on colloidal
  diffusion in corrugated channels}.
\newblock \emph{\bibinfo{journal}{Proc. Natl. Acad. Sci. U. S. A.}}
  \bibinfo{pages}{201707815} (\bibinfo{year}{2017}).

\bibitem{martens2013hydrodynamically}
\bibinfo{author}{Martens, S.}, \bibinfo{author}{Straube, A.},
  \bibinfo{author}{Schmid, G.}, \bibinfo{author}{Schimansky-Geier, L.} \&
  \bibinfo{author}{H{\"a}nggi, P.}
\newblock \bibinfo{title}{Hydrodynamically enforced entropic trapping of
  brownian particles}.
\newblock \emph{\bibinfo{journal}{Phys. Rev. Lett.}}
  \textbf{\bibinfo{volume}{110}}, \bibinfo{pages}{010601}
  (\bibinfo{year}{2013}).

\bibitem{kalinay2006corrections}
\bibinfo{author}{Kalinay, P.} \& \bibinfo{author}{Percus, J.}
\newblock \bibinfo{title}{Corrections to the fick-jacobs equation}.
\newblock \emph{\bibinfo{journal}{Phys. Rev. E}} \textbf{\bibinfo{volume}{74}},
  \bibinfo{pages}{041203} (\bibinfo{year}{2006}).

\bibitem{kalinay2017nonscaling}
\bibinfo{author}{Kalinay, P.}
\newblock \bibinfo{title}{Nonscaling calculation of the effective diffusion
  coefficient in periodic channels}.
\newblock \emph{\bibinfo{journal}{J. Chem. Phys.}}
  \textbf{\bibinfo{volume}{146}}, \bibinfo{pages}{034109}
  (\bibinfo{year}{2017}).

\bibitem{martens2011entropic}
\bibinfo{author}{Martens, S.}, \bibinfo{author}{Schmid, G.},
  \bibinfo{author}{Schimansky-Geier, L.} \& \bibinfo{author}{H{\"a}nggi, P.}
\newblock \bibinfo{title}{Entropic particle transport: Higher-order corrections
  to the fick-jacobs diffusion equation}.
\newblock \emph{\bibinfo{journal}{Phys. Rev. E}} \textbf{\bibinfo{volume}{83}},
  \bibinfo{pages}{051135} (\bibinfo{year}{2011}).

\bibitem{zwanzig1992diffusion}
\bibinfo{author}{Zwanzig, R.}
\newblock \bibinfo{title}{Diffusion past an entropy barrier}.
\newblock \emph{\bibinfo{journal}{J Phys. Chem.}}
  \textbf{\bibinfo{volume}{96}}, \bibinfo{pages}{3926--3930}
  (\bibinfo{year}{1992}).

\bibitem{mangeat2017dispersion}
\bibinfo{author}{Mangeat, M.}, \bibinfo{author}{Guerin, T.} \&
  \bibinfo{author}{Dean, D.~S.}
\newblock \bibinfo{title}{Dispersion in two dimensional channels—the
  Fick--Jacobs approximation revisited}.
\newblock \emph{\bibinfo{journal}{J. Stat. Mech.: Theor. Exp.}}
  \textbf{\bibinfo{volume}{2017}}, \bibinfo{pages}{123205}
  (\bibinfo{year}{2017}).

\bibitem{mangeat2017geometry}
\bibinfo{author}{Mangeat, M.}, \bibinfo{author}{Gu{\'e}rin, T.} \&
  \bibinfo{author}{Dean, D.~S.}
\newblock \bibinfo{title}{Geometry controlled dispersion in periodic corrugated
  channels}.
\newblock \emph{\bibinfo{journal}{Europhys. Lett.}}
  \textbf{\bibinfo{volume}{118}}, \bibinfo{pages}{40004}
  (\bibinfo{year}{2017}).

\bibitem{mangeat2018dispersion}
\bibinfo{author}{Mangeat, M.}, \bibinfo{author}{Gu{\'e}rin, T.} \&
  \bibinfo{author}{Dean, D.}
\newblock \bibinfo{title}{Dispersion in two-dimensional periodic channels with
  discontinuous profiles}.
\newblock \emph{\bibinfo{journal}{J. Chem. Phys.}}
  \textbf{\bibinfo{volume}{149}}, \bibinfo{pages}{124105}
  (\bibinfo{year}{2018}).


\bibitem{israelachvili2011intermolecular}
\bibinfo{author}{Israelachvili, J.~N.}
\newblock \emph{\bibinfo{title}{Intermolecular and surface forces, second edition}}
 (\bibinfo{publisher}{Academic press, London}
\bibinfo{year}{1991}).

\bibitem{brenner1961slow}
\bibinfo{author}{Brenner, H.}
\newblock \bibinfo{title}{The slow motion of a sphere through a viscous fluid
  towards a plane surface}.
\newblock \emph{\bibinfo{journal}{Chem. Eng. Sci.}}
  \textbf{\bibinfo{volume}{16}}, \bibinfo{pages}{242--251}
  (\bibinfo{year}{1961}).

\bibitem{bychuk1995anomalous}
\bibinfo{author}{Bychuk, O.~V.} \& \bibinfo{author}{O'Shaughnessy, B.}
\newblock \bibinfo{title}{Anomalous diffusion at liquid surfaces}.
\newblock \emph{\bibinfo{journal}{Phys. Rev. Lett.}}
  \textbf{\bibinfo{volume}{74}}, \bibinfo{pages}{1795} (\bibinfo{year}{1995}).

\bibitem{walder2011single}
\bibinfo{author}{Walder, R.}, \bibinfo{author}{Nelson, N.} \&
  \bibinfo{author}{Schwartz, D.~K.}
\newblock \bibinfo{title}{Single molecule observations of desorption-mediated
  diffusion at the solid-liquid interface}.
\newblock \emph{\bibinfo{journal}{Phys. Rev. Lett.}}
  \textbf{\bibinfo{volume}{107}}, \bibinfo{pages}{156102}
  (\bibinfo{year}{2011}).

\bibitem{skaug2013intermittent}
\bibinfo{author}{Skaug, M.~J.}, \bibinfo{author}{Mabry, J.} \&
  \bibinfo{author}{Schwartz, D.~K.}
\newblock \bibinfo{title}{Intermittent molecular hopping at the solid-liquid
  interface}.
\newblock \emph{\bibinfo{journal}{Phys. Rev. Lett.}}
  \textbf{\bibinfo{volume}{110}}, \bibinfo{pages}{256101}
  (\bibinfo{year}{2013}).

\bibitem{chee2016desorption}
\bibinfo{author}{Chee, S.~W.}, \bibinfo{author}{Baraissov, Z.},
  \bibinfo{author}{Loh, N.~D.}, \bibinfo{author}{Matsudaira, P.~T.} \&
  \bibinfo{author}{Mirsaidov, U.}
\newblock \bibinfo{title}{Desorption-mediated motion of nanoparticles at the
  liquid--solid interface}.
\newblock \emph{\bibinfo{journal}{J. Phys. Chem. C}}
  \textbf{\bibinfo{volume}{120}}, \bibinfo{pages}{20462--20470}
  (\bibinfo{year}{2016}).

\bibitem{wang2020non}
\bibinfo{author}{Wang, D.} \& \bibinfo{author}{Schwartz, D.~K.}
\newblock \bibinfo{title}{Non-brownian interfacial diffusion: Flying, hopping,
  and crawling}.
\newblock \emph{\bibinfo{journal}{J. Phys. Chem. C}}
  \textbf{\bibinfo{volume}{124}}, \bibinfo{pages}{19880--19891}
  (\bibinfo{year}{2020}).

\bibitem{morrin2020polyelectrolyte}
\bibinfo{author}{Morrin, G.~T.}, \bibinfo{author}{Kienle, D.~F.},
  \bibinfo{author}{Weltz, J.~S.}, \bibinfo{author}{Traeger, J.~C.} \&
  \bibinfo{author}{Schwartz, D.~K.}
\newblock \bibinfo{title}{Polyelectrolyte surface diffusion in a nanoslit
  geometry}.
\newblock \emph{\bibinfo{journal}{Macromolecules}}
  \textbf{\bibinfo{volume}{53}}, \bibinfo{pages}{4110--4120}
  (\bibinfo{year}{2020}).

\bibitem{benichou2010optimal}
\bibinfo{author}{B{\'e}nichou, O.}, \bibinfo{author}{Grebenkov, D.},
  \bibinfo{author}{Levitz, P.}, \bibinfo{author}{Loverdo, C.} \&
  \bibinfo{author}{Voituriez, R.}
\newblock \bibinfo{title}{Optimal reaction time for surface-mediated
  diffusion}.
\newblock \emph{\bibinfo{journal}{Phys. Rev. Lett.}}
  \textbf{\bibinfo{volume}{105}}, \bibinfo{pages}{150606}
  (\bibinfo{year}{2010}).

\bibitem{calandre2014accelerating}
\bibinfo{author}{Calandre, T.}, \bibinfo{author}{B{\'e}nichou, O.} \&
  \bibinfo{author}{Voituriez, R.}
\newblock \bibinfo{title}{Accelerating search kinetics by following
  boundaries}.
\newblock \emph{\bibinfo{journal}{Phys. Rev. Lett.}}
  \textbf{\bibinfo{volume}{112}}, \bibinfo{pages}{230601}
  (\bibinfo{year}{2014}).

\bibitem{rupprecht2012exact}
\bibinfo{author}{Rupprecht, J.-F.}, \bibinfo{author}{B{\'e}nichou, O.},
  \bibinfo{author}{Grebenkov, D.} \& \bibinfo{author}{Voituriez, R.}
\newblock \bibinfo{title}{Exact mean exit time for surface-mediated diffusion}.
\newblock \emph{\bibinfo{journal}{Phys. Rev. E}} \textbf{\bibinfo{volume}{86}},
  \bibinfo{pages}{041135} (\bibinfo{year}{2012}).

\bibitem{monserud2016interfacial}
\bibinfo{author}{Monserud, J.~H.} \& \bibinfo{author}{Schwartz, D.~K.}
\newblock \bibinfo{title}{Interfacial molecular searching using forager
  dynamics}.
\newblock \emph{\bibinfo{journal}{Phys. Rev. Lett.}}
  \textbf{\bibinfo{volume}{116}}, \bibinfo{pages}{098303}
  (\bibinfo{year}{2016}).

\bibitem{von1989facilitated}
\bibinfo{author}{von Hippel, P.~H.} \& \bibinfo{author}{Berg, O.~G.}
\newblock \bibinfo{title}{Facilitated target location in biological systems}.
\newblock \emph{\bibinfo{journal}{J. Biol. Chem.}}
  \textbf{\bibinfo{volume}{264}}, \bibinfo{pages}{675--678}
  (\bibinfo{year}{1989}).

\bibitem{coppey2004kinetics}
\bibinfo{author}{Coppey, M.}, \bibinfo{author}{B{\'e}nichou, O.},
  \bibinfo{author}{Voituriez, R.} \& \bibinfo{author}{Moreau, M.}
\newblock \bibinfo{title}{Kinetics of target site localization of a protein on
  dna: a stochastic approach}.
\newblock \emph{\bibinfo{journal}{Biophys. J.}} \textbf{\bibinfo{volume}{87}},
  \bibinfo{pages}{1640--1649} (\bibinfo{year}{2004}).

\bibitem{berg1981diffusion}
\bibinfo{author}{Berg, O.~G.}, \bibinfo{author}{Winter, R.~B.} \&
  \bibinfo{author}{Von~Hippel, P.~H.}
\newblock \bibinfo{title}{Diffusion-driven mechanisms of protein translocation
  on nucleic acids. 1. models and theory}.
\newblock \emph{\bibinfo{journal}{Biochemistry}} \textbf{\bibinfo{volume}{20}},
  \bibinfo{pages}{6929--6948} (\bibinfo{year}{1981}).

\bibitem{putzel2014nonmonotonic}
\bibinfo{author}{Putzel, G.~G.}, \bibinfo{author}{Tagliazucchi, M.} \&
  \bibinfo{author}{Szleifer, I.}
\newblock \bibinfo{title}{Nonmonotonic diffusion of particles among larger
  attractive crowding spheres}.
\newblock \emph{\bibinfo{journal}{Phys. Rev. Lett.}}
  \textbf{\bibinfo{volume}{113}}, \bibinfo{pages}{138302}
  (\bibinfo{year}{2014}).

\bibitem{levesque2012taylor}
\bibinfo{author}{Levesque, M.}, \bibinfo{author}{B{\'e}nichou, O.},
  \bibinfo{author}{Voituriez, R.} \& \bibinfo{author}{Rotenberg, B.}
\newblock \bibinfo{title}{Taylor dispersion with adsorption and desorption}.
\newblock \emph{\bibinfo{journal}{Phys. Rev. E}} \textbf{\bibinfo{volume}{86}},
  \bibinfo{pages}{036316} (\bibinfo{year}{2012}).

\bibitem{berezhkovskii2013aris}
\bibinfo{author}{Berezhkovskii, A.~M.} \& \bibinfo{author}{Skvortsov, A.~T.}
\newblock \bibinfo{title}{Aris-taylor dispersion with drift and diffusion of
  particles on the tube wall}.
\newblock \emph{\bibinfo{journal}{J. Chem. Phys.}}
  \textbf{\bibinfo{volume}{139}}, \bibinfo{pages}{084101}
  (\bibinfo{year}{2013}).

\bibitem{quintard1994convection}
\bibinfo{author}{Quintard, M.} \& \bibinfo{author}{Whitaker, S.}
\newblock \bibinfo{title}{Convection, dispersion, and interfacial transport of
  contaminants: Homogeneous porous media}.
\newblock \emph{\bibinfo{journal}{Adv. Water. Ress.}}
  \textbf{\bibinfo{volume}{17}}, \bibinfo{pages}{221--239}
  (\bibinfo{year}{1994}).

\bibitem{edwards1993dispersion}
\bibinfo{author}{Edwards, D.~A.}, \bibinfo{author}{Shapiro, M.} \&
  \bibinfo{author}{Brenner, H.}
\newblock \bibinfo{title}{Dispersion and reaction in two-dimensional model
  porous media}.
\newblock \emph{\bibinfo{journal}{Physics of Fluids A: Fluid Dynamics}}
  \textbf{\bibinfo{volume}{5}}, \bibinfo{pages}{837--848}
  (\bibinfo{year}{1993}).

\bibitem{mangeat2020effective}
\bibinfo{author}{Mangeat, M.} \bibinfo{author}{Guérin, T.} \& \bibinfo{author}{Dean, D.S.} 
\newblock \bibinfo{title}{Effective diffusivity of Brownian particles in a two dimensional square lattice of hard disks}.
\newblock \emph{\bibinfo{journal}{J. Chem. Phys.}}
  \textbf{\bibinfo{volume}{152}}, \bibinfo{pages}{234109}
  (\bibinfo{year}{2020}).

\bibitem{kalnin2002calculations}
\bibinfo{author}{Kalnin, J.R.}, \bibinfo{author}{Kotomin, E.A.}, \& \bibinfo{author}{Maier, J.}
\newblock \bibinfo{title}{Calculations of the effective diffusion coefficient for inhomogeneous media}
\newblock \emph{\bibinfo{journal}{Journal of physics and chemistry of solids}},
  \textbf{\bibinfo{volume}{63}}
   \bibinfo{pages}{449--456}
     (\bibinfo{year}{2002}).

\bibitem{MaxwellBook}
\bibinfo{author}{Maxwell, J. C.}, 
\newblock \emph{\bibinfo{title}{Electricity and Magnetism}}
  (\bibinfo{publisher}{Clarendon Press, Oxford}, \bibinfo{year}{1873}).

\bibitem{lebenhaft1979diffusion}
\bibinfo{author}{Lebenhaft, J.R.}, \bibinfo{author}{Kapral, R.}, \& \bibinfo{author}{Maier, J.}
\newblock \bibinfo{title}{Diffusion-controlled processes among partially absorbing stationary sinks}
\newblock \emph{\bibinfo{journal}{J. Stat. Phys.}},
  \textbf{\bibinfo{volume}{20}}
   \bibinfo{pages}{25-56}
     (\bibinfo{year}{1979}).


\bibitem{grebenkov2019imperfect}
\bibinfo{author}{Grebenkov, D.~S.}
\newblock \bibinfo{title}{Imperfect diffusion-controlled reactions}.
\newblock \emph{\bibinfo{journal}{Chemical Kinetics: Beyond the Textbook}}
  \bibinfo{pages}{191--219} (\bibinfo{year}{2019}).

\bibitem{AlexandreGeneralized2021}
\bibinfo{author}{Alexandre, A.}, \bibinfo{author}{Guérin, T.} \&
  \bibinfo{author}{Dean, D.~S.}
\newblock \bibinfo{title}{Generalized taylor dispersion for translationally
  invariant microfluidic systems}.
\newblock \emph{\bibinfo{journal}{Phys. Fluids}} \textbf{\bibinfo{volume}{33}},
  \bibinfo{pages}{082004} (\bibinfo{year}{2021}).

%\bibitem{SINotes}
%See Supplemental Material at [http://...], which  includes   Refs.~\cite{singer2008partially,ward1993,mamode2014fundamental,lifson1962self}

\bibitem{singer2008partially}
\bibinfo{author}{Singer, A.}, \bibinfo{author}{Schuss, Z.},
  \bibinfo{author}{Osipov, A.} \& \bibinfo{author}{Holcman, D.}
\newblock \bibinfo{title}{Partially reflected diffusion}.
\newblock \emph{\bibinfo{journal}{SIAM J. Appl. Math.}}
  \textbf{\bibinfo{volume}{68}}, \bibinfo{pages}{844--868}
  (\bibinfo{year}{2008}).

\bibitem{ward1993}
\bibinfo{author}{Ward, M.~J.} \& \bibinfo{author}{Keller, J.~B.}
\newblock \bibinfo{title}{Strong localized perturbations of eigenvalue
  problems}.
\newblock \emph{\bibinfo{journal}{SIAM J. Appl. Math.}}
  \textbf{\bibinfo{volume}{53}}, \bibinfo{pages}{770--798}
  (\bibinfo{year}{1993}).

\bibitem{mamode2014fundamental}
\bibinfo{author}{Mamode, M.}
\newblock \bibinfo{title}{Fundamental solution of the laplacian on flat tori
  and boundary value problems for the planar poisson equation in rectangles}.
\newblock \emph{\bibinfo{journal}{Boundary Value Problems}}
  \textbf{\bibinfo{volume}{2014}}, \bibinfo{pages}{1--9}
  (\bibinfo{year}{2014}).

 
\bibitem{lifson1962self}
\bibinfo{author}{Lifson, S.} \& \bibinfo{author}{Jackson, J.~L.}
\newblock \bibinfo{title}{On the self-diffusion of ions in a polyelectrolyte
  solution}.
\newblock \emph{\bibinfo{journal}{J. Chem. Phys.}}
  \textbf{\bibinfo{volume}{36}}, \bibinfo{pages}{2410--2414}
  (\bibinfo{year}{1962}).

\bibitem{Guerin2015Kubo}
\bibinfo{author}{Gu\'erin, T.} \& \bibinfo{author}{Dean, D.~S.}
\newblock \bibinfo{title}{Kubo formulas for dispersion in heterogeneous
  periodic nonequilibrium systems}.
\newblock \emph{\bibinfo{journal}{Phys. Rev. E}} \textbf{\bibinfo{volume}{92}},
  \bibinfo{pages}{062103} (\bibinfo{year}{2015}).

\bibitem{guerin2015}
\bibinfo{author}{Gu\'erin, T.} \& \bibinfo{author}{Dean, D.~S.}
\newblock \bibinfo{title}{Force-induced dispersion in heterogeneous media}.
\newblock \emph{\bibinfo{journal}{Phys. Rev. Lett.}}
  \textbf{\bibinfo{volume}{115}}, \bibinfo{pages}{020601}
  (\bibinfo{year}{2015}).

\bibitem{edward1995diffusion}
\bibinfo{author}{Edwards, D.~A.} \& \bibinfo{author}{Davis, A.}
\newblock \bibinfo{title}{Diffusion and convective dispersion through arrays of
  spheres with surface adsorption, diffusion, and unequal solute partitioning}.
\newblock \emph{\bibinfo{journal}{Chemical engineering science}}
  \textbf{\bibinfo{volume}{50}}, \bibinfo{pages}{1441--1454}
  (\bibinfo{year}{1995}).

\end{thebibliography}

\end{document}